\documentclass[aps,prd,superscriptaddress,amsmath,groupedaddress,twocolumn,nofootinbib]{revtex4}

\usepackage{color,hhline,psfrag,rotating,amssymb}
\usepackage{slashed}
\usepackage{dcolumn}

\begin{document}
\title{$B \rightarrow \pi \ell \nu$ at zero recoil from lattice QCD with physical $u/d$ quarks}
\author{B. Colquhoun}
\affiliation{SUPA, School of Physics and Astronomy, University of Glasgow, Glasgow, G12 8QQ, UK}
\author{R. J. Dowdall}
\affiliation{DAMTP, University of Cambridge, Wilberforce Road, Cambridge, CB3 0WA, UK}
\author{J. Koponen}
\affiliation{SUPA, School of Physics and Astronomy, University of Glasgow, Glasgow, G12 8QQ, UK}
\author{C. T. H. Davies}
\email[]{christine.davies@glasgow.ac.uk}
\affiliation{SUPA, School of Physics and Astronomy, University of Glasgow, Glasgow, G12 8QQ, UK}
\author{G. P. Lepage}
\affiliation{Laboratory of Elementary-Particle Physics, Cornell University, Ithaca, New York 14853, USA}

\collaboration{HPQCD collaboration}
\homepage{http://www.physics.gla.ac.uk/HPQCD}
\noaffiliation

\date{\today}

\begin{abstract}
The exclusive semileptonic decay $B \rightarrow \pi \ell \nu$ is a 
key process for the determination of the Cabibbo-Kobayashi-Maskawa 
matrix element $V_{ub}$ from the comparison of experimental rates as 
a function of $q^2$ with 
theoretically determined form factors. 
The sensitivity of the form factors to the $u/d$ quark mass has meant 
significant systematic uncertainties in lattice QCD calculations at 
unphysically heavy pion masses. 
Here we give the first lattice QCD calculations of this process for 
u/d quark masses going down to their physical values, 
calculating the $f_0$ form factor at zero 
recoil to 3\%. 
We are able to resolve a long-standing controversy by 
showing that the soft-pion theorem result $f_0(q^2_{max}) = f_B/f_{\pi}$ 
does hold as $m_{\pi} \rightarrow 0$. We use the Highly Improved Staggered 
Quark formalism for the light quarks and show that 
staggered chiral perturbation theory for the $m_{\pi}$ dependence 
is almost identical to continuum chiral perturbation theory for $f_0$, 
$f_B$ and $f_{\pi}$. 
We also give results for other 
processes such as $B_s \rightarrow K \ell \nu$.  

\end{abstract}


\maketitle

\section{Introduction}
\label{sec:intro}
The exclusive semileptonic process, $B \rightarrow \pi \ell \nu$,
is a key one for flavour physics 
because it gives access to the Cabibbo-Kobayashi-Maskawa matrix 
element $V_{ub}$ from a process involving two `gold-plated' (stable in QCD) 
mesons, $B$ and $\pi$. $V_{ub}$ is determined by comparing the 
experimental rate for the process to that determined from   
theoretical calculations of hadronic parameters known as 
form factors which are functions of the squared 4-momentum transfer, 
$q^2$, between the $B$ and the $\pi$. The form factors encapsulate 
the information about how likely it is for a $\pi$ meson with a specific 
momentum to form when a $b$ quark inside a $B$ meson changes into 
a $u$ quark with emission of a $W$ boson. If the calculations of 
the form factors are done in lattice QCD then the full effect 
of QCD interactions that keep the quarks bound inside the mesons
is taken into account. 

Lattice QCD calculations for this process are 
particularly difficult, however, because the results are sensitive to 
the mass of the $u/d$ quarks that form the $\pi$ meson.   
Existing lattice QCD calculations have used $u/d$ quarks 
that have heavier masses than in the real world and results then 
have to be 
extrapolated to the physical 
point. The value of the $u/d$ quark masses, and hence the $\pi$ mass, 
also affects the $q^2$ value for a given $\pi$ spatial momentum 
and, since the form factors are strongly varying functions of $q^2$, 
this gives further dependence on the quark masses.
For recent lattice QCD results for $B \rightarrow \pi \ell \nu$ form 
factors see~\cite{FNALbtopi2015, RBCbtopi2015}.

Lattice QCD calculations for the form factors are most reliable 
close to the `zero recoil' point where the $\pi$ has small 
spatial momentum (and hence relatively small discretisation 
errors). This corresponds to being close to the maximum value 
of $q^2$, $(m_B-m_{\pi})^2$. In this regime it is possible to 
use soft pion relations coupled to Heavy Quark Effective Theory 
to derive expectations for the functional dependence of 
form factors on the heavy quark mass, $m_b$, and on the $\pi$ meson 
mass. Since these relationships come from a well understood 
theoretical framework it is important to test them against 
lattice QCD results. 

It was apparent in quenched lattice QCD calculations many years 
ago that there were large deviations between lattice 
results and the expected dependence on $m_b$ and $m_{\pi}$;
see~\cite{Onogi:1997kb, Hashimoto:1999bk} for a review. 
Relatively heavy 
$u/d$ quark masses were used in these early calculations, often 
close to the $s$ quark mass. 

Here we revisit this 
issue with results from lattice QCD that include the full 
effect of sea quarks ($u$, $d$, $s$ and $c$) but, more 
importantly, include $u$ and $d$ quarks with masses at 
their very small physical values for the first time. We focus on the 
soft pion relation which gives the scalar form factor $f_0$ 
at the zero recoil point in terms of the ratio of 
$B$ and $\pi$ decay constants as $m_{\pi} \rightarrow 0$~\cite{Dominguez:1990mi, Wise:1992hn, Burdman:1992gh, Wolfenstein:1992xh}: 
\begin{equation}
\label{eq:softpion}
f_0(q^2_{max}) = \frac{f_B}{f_{\pi}}.
\end{equation}  
This relationship has been shown to hold through $\mathcal{O}(\Lambda/m_h)$
in an expansion in powers of the inverse heavy quark mass, 
$m_h$~\cite{Burdman:1993es}, and we therefore expect it to work at 
the 1\% level for $b$ quarks. A test of this in lattice 
QCD calculations can then provide important confirmation of how 
well we understand the $u/d$ quark mass dependence for a process
$B \rightarrow \pi \ell \nu$ for which this is critical. 

Here we show that indeed the soft-pion 
relation does emerge from lattice QCD calculations as $m_{\pi} \rightarrow 0$ 
for a fixed heavy quark mass tuned to that of the $b$.
This calculation builds on the very accurate results for 
$f_B$ and $f_{\pi}$ that we have been able to obtain 
at physical $u/d$ quark masses~\cite{DowdallBdecay, Dowdallfkpi} 
and benefits from the fact that the renormalisation factors of 
the temporal vector and temporal axial-vector heavy-light 
currents are the same for our light quark formalism and so 
renormalisation uncertainties are minimised.

In addition we give a precise result for $f_0(q^2_{max})$ 
for physical $u/d$ quark masses (where corrections to the soft-pion relation 
are substantial). This is a useful comparison point 
for lattice QCD calculations that are done with unphysical $u/d$ 
quark masses, as a test of extrapolations in the $u/d$ quark 
mass. 
Although $f_0$ itself is not accessible in experiment, lattice 
QCD calculations of it should agree, and it is important to test this 
using different discretisations of QCD. $f_0(q^2_{max})$ is 
the most accurately determined form factor for a lattice QCD 
calculation of $B \rightarrow \pi$ decay and so a good number 
for calibration. 

For comparison we also give results for $f_0(q^2_{max})$
for $B_s \rightarrow K \ell \nu$ decay, which is another 
physical process, and for $B_s \rightarrow \eta_s \ell \nu$,
which does not occur in the real world. The latter decay is again 
useful for comparison between lattice QCD calculations.  

The paper is laid out as follows: In Section~\ref{sec:latt} we 
describe the lattice calculation of the correlation functions needed 
to extract the scalar form factor and its ratio to 
$f_B/f_{\pi}$. We use the Highly Improved Staggered 
Quark (HISQ) action for the light quarks and improved NonRelativistic 
QCD (NRQCD) for the $b$ quark. In Section~\ref{sec:results} we give the 
results and determine the ratio both at the physical 
$m_{\pi}$ and as $m_{\pi} \rightarrow 0$. 
Section~\ref{sec:discussion} includes 
a comparison of our values with those 
from lattice QCD calculations 
that extrapolated results from heavier than physical values 
of $m_{\pi}$ and Section~\ref{sec:conclusions} 
provides our conclusions. In Appendix~\ref{appendix:schipt} 
we discuss the staggered quark chiral perturbation 
theory that we use for $f_B$, $f_{\pi}$ and $f_0$ and  
show that it is in fact very continuum-like 
in its approach to $m_{\pi}=0$. 

\section{Lattice calculation}
\label{sec:latt}
We use ensembles of lattice gluon 
configurations provided by the MILC collaboration~\cite{milchisq} 
at three values of the lattice 
spacing, $a \approx$  0.15 fm, 0.12 fm and 0.09 fm. 
The configurations include the effect of $u$, $d$, $s$ and $c$ 
quarks in the sea using the HISQ 
formalism~\cite{hisqdef} and a gluon action improved through 
$\mathcal{O}(\alpha_sa^2)$~\cite{Hart:2008sq}. These then give
significant improvements in the control of systematic 
errors from finite lattice spacing and light 
quark mass effects over earlier configurations. 

\begin{table}
\begin{tabular}{llllllll}
\hline
\hline
Set &  $a$/fm & $am_{l}$ & $am_{s}$ & $am_{c}$ & $L_s/a$ & $L_t/a$ & $N_{\mathrm{cfg}}$ \\
\hline
1 & 0.1474(5)(14) & 0.013 & 0.065 & 0.838 & 16 & 48 & 1020 \\
2 & 0.1463(3)(14) & 0.0064 & 0.064 & 0.828 & 24 & 48 & 1000 \\
3 & 0.1450(3)(14) & 0.00235 & 0.0647 & 0.831 & 32 & 48 & 1000 \\
\hline
4 & 0.1219(2)(9) & 0.0102 & 0.0509 & 0.635 & 24 & 64 & 1052 \\
5 & 0.1195(3)(9) & 0.00507 & 0.0507 & 0.628 & 32 & 64 & 1000 \\
6 & 0.1189(2)(9) & 0.00184 & 0.0507 & 0.628 & 48 & 64 & 1000 \\
\hline
7 & 0.0884(3)(5) & 0.0074 & 0.037 & 0.440 & 32 & 96 & 1008 \\
8 & 0.0873(2)(5) & 0.0012 & 0.0363 & 0.432 & 64 & 96 & 620 \\
\hline
\hline
\end{tabular}
\caption{ Details of gluon field configurations used in this 
calculation~\cite{milchisq}. $a$ is the lattice spacing, fixed from the mass 
difference between the $\Upsilon^{\prime}$ and $\Upsilon$ in~\cite{dowdallr1}.
The first error is from statistics and the second from NRQCD 
systematics in that determination and from experiment. 
Sets 1, 2 and 3 are `very coarse', sets 4, 5 and 6 are `coarse' and 
sets 7 and 8 are `fine'. 
$am_l$, $am_s$ and $am_c$ are the light ($u$ and 
$d$ are taken to have the same mass), strange and charm sea quark 
masses.
Sets 3, 6 and 8 have $m_l$ at close to its physical value. 
$L_s/a$ and $L_t/a$ are the number of lattice sites in 
the spatial and temporal directions respectively and $N_{cfg}$ is 
the number of configurations in the ensemble. We calculate 
propagators from 16 time sources on each ensemble (4 on set 8) 
to increase statistics. 
}
\label{tab:params}
\end{table}

We work at three different values of the $u/d$ quark masses (which 
are taken to be degenerate and will be referred to as light ($l$) 
quarks) in the sea. 
These correspond to approximately one-fifth and one-tenth of the $s$ quark mass 
and the physical average $u/d$ quark mass ($m_s/27.5$~\cite{pdg}). 
The lattice spacing on these configurations is determined for 
this calculation from 
the mass difference between the $\Upsilon^{\prime}$ and 
the $\Upsilon$~\cite{dowdallr1}. 
Table~\ref{tab:params} lists the 
parameters of the ensembles. 

On these configurations we calculate $l$ and $s$ quark propagators 
using the HISQ action. The $l$ quarks are taken to have the same 
mass as is used in the sea. For the $s$ quarks we retune the 
valence mass to be closer to the physical $s$ quark mass and so 
it is slightly different from that in the sea~\cite{dowdallr1}. Values 
are given in Table~\ref{tab:mbc}. We also 
calculate $b$ quark propagators using the improved NRQCD~\cite{nrqcd} 
action developed in~\cite{dowdallr1, nrqcdimp}. 

\begin{table}
\begin{tabular}{lllllll}
\hline
\hline
Set &  $am_s^{\rm val}$ & $am_b$ &  $u_{0L}$ & $c_1,c_6$ & $c_5$ & $c_4$ \\
\hline
1 & 0.0641 & 3.297 & 0.8195 & 1.36 & 1.21 & 1.22 \\
2 & 0.0636 & 3.263 & 0.82015 & 1.36 & 1.21 & 1.22 \\
3 & 0.0628 & 3.25  & 0.8195 & 1.36 & 1.21 & 1.22 \\
\hline
4 & 0.0522 & 2.66 & 0.8340 & 1.31 & 1.16 & 1.20 \\
5 & 0.0505 & 2.62 & 0.8349 & 1.31 & 1.16 & 1.20 \\
6 & 0.0507 & 2.62 & 0.8341 & 1.31 & 1.16 & 1.20 \\
\hline
7 & 0.0364 & 1.91 & 0.8525 & 1.21 & 1.12 & 1.16 \\
8 & 0.0360 & 1.89 & 0.8518 & 1.21 & 1.12 & 1.16 \\
\hline
\hline
\end{tabular}
\caption{ Summary of the valence $s$ quark mass and valence $b$ quark mass 
and other action parameters 
for the NRQCD action on the different ensembles of Table~\ref{tab:params}. 
The $s$ and $b$ quark masses in lattice units (columns 2 and 3) were tuned 
in~\cite{dowdallr1,Colquhoun:2014ica}. 
Column 4 gives the parameter $u_{0L}$ used for `tadpole-improving' 
the gluon field~\cite{dowdallr1, DowdallBdecay} and columns 5, 6 and 7 give the 
coefficients of kinetic and chromomagnetic terms used in the 
NRQCD action. $c_1$ ($c_6$ has the same value), $c_5$ and $c_4$ 
are correct through 
$\mathcal{O}(\alpha_s)$~\cite{dowdallr1, nrqcdimp}. 
}
\label{tab:mbc}
\end{table}

The NRQCD Hamiltonian we use is given by~\cite{nrqcd}:
 \begin{eqnarray}
\label{eq:evolution}
 e^{-aH} &=& \left(1-\frac{a\delta H}{2}\right)\left(1-\frac{aH_0}{2n_h}\right)^{n_h} U_t^{\dag} \nonumber \\
&& \times \left(1-\frac{aH_0}{2n_h}\right)^{n_h}\left(1-\frac{a\delta H}{2}\right) 
\end{eqnarray}
with
 \begin{eqnarray}
 aH_0 &=& - \frac{\Delta^{(2)}}{2 am_b}, \nonumber \\
a\delta H
&=& - c_1 \frac{(\Delta^{(2)})^2}{8( am_b)^3}
            + c_2 \frac{i}{8(am_b)^2}\left(\bf{\nabla}\cdot\tilde{\bf{E}}\right. -
\left.\tilde{\bf{E}}\cdot\bf{\nabla}\right) \nonumber \\
& & - c_3 \frac{1}{8(am_b)^2} \bf{\sigma}\cdot\left(\tilde{\bf{\nabla}}\times\tilde{\bf{E}}\right. -
\left.\tilde{\bf{E}}\times\tilde{\bf{\nabla}}\right) \nonumber \\
 & & - c_4 \frac{1}{2 am_b}\,{\bf{\sigma}}\cdot\tilde{\bf{B}}  
  + c_5 \frac{\Delta^{(4)}}{24 am_b} \nonumber \\
 & & -  c_6 \frac{(\Delta^{(2)})^2}{16n_h(am_b)^2} .
\label{eq:deltaH}
\end{eqnarray}
Here $\nabla$ is the symmetric lattice derivative and $\Delta^{(2)}$ and 
$\Delta^{(4)}$ the lattice discretization of the continuum $\sum_iD_i^2$ and 
$\sum_iD_i^4$ respectively. $am_b$ is the bare $b$ quark mass in units 
of the lattice spacing. $n_h$ is a stability parameter set equal to 
4 here. The gluon field is tadpole-improved, 
which means dividing all the links, $U_{\mu}(x)$ 
by a tadpole-parameter, $u_0$, before constructing covariant 
derivatives or chromo-electric or magnetic fields. 
For $u_0$ we took the mean trace of the gluon field 
in Landau gauge, $u_{0L}$~\cite{dowdallr1}. 
$\bf \tilde{E}$ and $\bf \tilde{B}$ are the chromoelectric 
and chromomagnetic fields calculated from an improved clover term~\cite{gray}.
They are made anti-hermitian 
but not explicitly traceless, to match the perturbative calculations 
done using this action.  

Given the NRQCD action above, 
the heavy quark propagator is readily calculated from 
a simple lattice time evolution equation~\cite{nrqcd} which is 
numerically very fast. 
The heavy quark propagator is given by:
\begin{equation}
G_b({\bf x},t+1) = e^{-aH}G_b({\bf x},t) 
\label{eq:evol}
\end{equation}
with starting condition:
\begin{equation}
G_b({\bf x},0) = \phi({\bf x})\Gamma({\bf x}).
\label{eq:nrqcdstart}
\end{equation}
Here $\Gamma$ is a matrix in (2-component) spin space 
and 
$\phi({\bf x})$ is a 
function of spatial position, to be discussed below. 
We can use such a function because we fix the gluon field 
configurations to Coulomb gauge. 

The terms in $\delta H$ in eq.~(\ref{eq:deltaH}) have 
coefficients $c_i$ whose values can be
fixed from matching lattice NRQCD to full QCD perturbatively, 
 giving the $c_i$ the 
expansion $1 + c^{(1)}_i\alpha_s + \mathcal{O}(\alpha_s^2)$. 
Here we include $\mathcal{O}(\alpha_s)$ corrections to the coefficients 
of the sub-leading kinetic terms, $c_1$, $c_5$ and $c_6$, 
and the chromomagnetic term, $c_4$~\cite{dowdallr1, nrqcdimp}. 
The $b$ quark mass parameter, $am_b$, is nonperturbatively tuned to the correct 
value 
by calculating the spin-average of the `kinetic masses' 
of the $\Upsilon$ and $\eta_b$ as described in~\cite{dowdallr1}.  
We are able to do this to 1\%, limited by the accuracy of the determination of the 
lattice spacing to convert the kinetic mass to physical units. 
The values used for $am_b$, $u_{0L}$ and $c_i$ on the different ensembles 
are given in Table~\ref{tab:mbc}. 

We use the same NRQCD action for both heavyonium and for 
heavy-light meson 
calculations since it is accurate for both. 
For heavy-light calculations, which concern us here, the 
power counting in the heavy-quark velocity is 
equivalent to power-counting in inverse powers of the heavy quark 
mass. 
The NRQCD action above is then fully improved through 
$\mathcal{O}(\alpha_s\Lambda/m_b)$. 
It has already been used for accurate calculations 
of the $\Upsilon$ spectrum and properties~\cite{dowdallr1, Daldrop, Dowdallhyp, Colquhoun:2014ica} 
and $B$, $B_s$ and $B_c$ meson masses~\cite{rachelBmass} and decay 
constants~\cite{DowdallBdecay, Colquhoun:2015oha}, including those of 
their vector partners. 

\begin{table}
\begin{tabular}{lll}
\hline
\hline
Set &  $a_{sm}$ & $T$ \\
\hline
1,2 & 2.0,4.0 & 10,13,16 \\
3  & 2.0,4.0 & 9,12,15 \\
\hline
4 & 2.5,5.0 & 14,19,24 \\
5,6 & 2.0,4.0 & 14,19,24 \\
\hline
7 & 3.425,6.85 & 19,24,29 \\
8 & 3.425,6.85 & 20,27,34 \\
\hline
\hline
\end{tabular}
\caption{ Summary of the smearing radii of the smearing 
functions used for the $b$ quark propagators on the different 
sets of configurations in Table~\ref{tab:params}. 
Column 3 gives the different 
$T$ values used for the creation time-slice for the $B$ meson.  
}
\label{tab:smear}
\end{table}

For the calculation of the $B \rightarrow \pi \ell \nu$ form factor we 
need (Goldstone) $\pi$ meson correlation functions, $B$ meson correlation functions 
and `3-point' correlation functions that connect the $B$ meson to the 
$\pi$. Here we work with $B$ and $\pi$ mesons at rest. The $\pi$ meson 
correlators were calculated in~\cite{Dowdallfkpi} and are simply 
given by:
\begin{equation}
C_{\pi}(t) = \frac{1}{4} \left< \sum_{\vec{x}} \mathrm{tr} \left| g(\vec{x},t) \right|^2 \right>
\label{eq:picorr}
\end{equation}
where $g(\vec{x},t)$ is a staggered quark propagator with source at $t=0$, 
$\mathrm{tr}$ denotes a trace over color indices and the angle brackets 
denote an average over gluon field configurations in the ensemble. 
We use random wall 
sources to improve statistical errors. The factor of 1/4 is needed to account
for the number of staggered quark `tastes' because the correlator is 
 a staggered quark loop. 
Fitting these `2-point' correlators to the standard multi-exponential form 
\begin{equation}
C_{\pi}(t) = \sum_{k=0}^{n_{exp}-1} a_{\pi,k}^2 \left( e^{-E_{\pi,k}t} + e^{-E_{\pi,k}(T-t)} \right)
\label{eq:pifit}
\end{equation}
allows the extraction of the ground-state $\pi$ mass ($\equiv E_{\pi,0}$). The amplitude 
$a_{\pi,0}$ gives $\langle 0 | P | \pi \rangle/(\sqrt{2m_{\pi}})$ where $P$ is the 
pseudoscalar density. Using the PCAC relation we can convert this to $f_{\pi}$:
\begin{equation}
f_{\pi} = 2m_l a_{\pi,0} \sqrt{(2/E_{\pi,0}^3)}
\label{eq:fpi}
\end{equation}
and note that $f_{\pi}$ is absolutely normalised here. 
Results for $m_{\pi}$ and $f_{\pi}$ are given in~\cite{Dowdallfkpi}. Statistical 
errors below 0.1\% are obtained. 
 
Staggered quarks have numerical efficiency advantages as a result of 
having no spin degree of freedom. A 4-component naive quark propagator, $\tilde{g}$, 
can be simply obtained from a staggered quark propagator, $g$, from point $x$ to 
point $y$ by 
reversing the staggering transformation: 
\begin{eqnarray}
\tilde{g}(x,y) &=& g(x,y)\Omega(x)\Omega^{\dag}(y) \\
\Omega(x) &=& \prod_{\mu=1}^4 (\gamma_{\mu})^{x_{\mu}}. \nonumber
\label{eq:stagg}
\end{eqnarray}
We can then use $\tilde{g}$ straightforwardly in combination with 
other propagators that carry a spin component~\cite{Wingate}, for example an 
NRQCD $b$ quark propagator. $B$ meson correlators at zero spatial 
momentum were made in this 
way in~\cite{DowdallBdecay}. NRQCD $b$ quark and staggered light 
quark propagators are simply combined to make a pseudoscalar meson 
(with operator $\overline{\Psi}_b(x)\gamma_5 \Psi_l(x)$) as:
\begin{equation}
C_{B}(t) = \left< \sum_{\vec{x}} \mathrm{tr}[ \mathrm{Tr}\Omega^{\dag}(x) G_b(\vec{x},t) ] g^{\dag}(\vec{x},t)  \right> .
\label{eq:Bcorr}
\end{equation}
Here $\mathrm{Tr}$ is a spin trace. In this equation both $G_b$ and 
$g$ are calculated from a simple delta function source at the origin (and $\Omega(0)=1$). 

To improve our statistical precision we use a random wall source for 
both propagators which adds the technical complication 
that $\Omega^{\dag}(\vec{x},t=0)$ must be 
used, along with the U(1) random noise field, as the source for the NRQCD propagator~\cite{gregory}. In addition 
we use 3 different smearing functions for the source of the $b$ quark 
in eq.~(\ref{eq:nrqcdstart}): a delta function and two exponentials with 
different radii, $a_{sm}$, given in Table~\ref{tab:smear}.  
Thus, in eq.~(\ref{eq:nrqcdstart}), $\Gamma=\Omega(\vec{x},t=0)$ and, 
for the smeared case, 
\begin{equation}
\phi(\vec{x}) = \sum_{\vec{y}} e^{(-|\vec{x}-\vec{y}|/a_{sm})}\eta(\vec{y})
\end{equation}
with $\eta(\vec{y})$ a random field from U(1), and a 3-vector in colour space. 
For the staggered quark propagator the source is simply $\eta(\vec{x})$. 

We obtain a $3\times 3$ matrix of correlation functions from 
the use of the 3 different smearings for the $b$ quark at source and sink. 
This can be fit to a multi-exponential form as for the $\pi$ meson, except 
that the correlator has a simple exponential form in time rather than a cosh 
because NRQCD $b$ quarks propagate in one direction in time only. 
To improve statistics we average over forward-in-time and backward-in-time 
directions.
The fit enables us to 
extract $B$ meson energies (these are not 
equal to the meson masses 
because of the NRQCD energy offset) and amplitudes that depend on 
the smearing function used~\cite{rachelBmass}.  We use a fit function
\begin{eqnarray}
C_B(t) &=& \sum_{m=0}^{m_{exp}-1} c(\phi_{sc},m)c^*(\phi_{sk},m)e^{-E_mt} \\
&-& (-1)^t\sum_{m^{\prime}=1}^{m_{exp}-1} d(\phi_{sc},m^{\prime})d^*(\phi_{sk},m^{\prime})e^{-E^{\prime}_{m^{\prime}}t}, \nonumber
\label{eq:corrfit}
\end{eqnarray}
where $sc$ and $sk$ denote source and sink respectively. The second line captures
the presence of `oscillating' opposite parity states in the correlator that 
are a result of using a staggered light quark. 
Having multiple smearing functions improves the fit significantly because 
they enhance the signal for the ground-state at small $t$ values which 
counteracts the relatively poor signal-noise in $B$ correlators at large $t$. 

The decay constant for the $B$ meson is defined from the 
matrix element of the continuum temporal axial current between 
the vacuum and a $B$ meson at rest: 
\begin{equation}
f_Bm_B = \langle 0 | A_0 | B(\vec{p}=0) \rangle . 
\end{equation}
To determine this accurately in our lattice QCD calculations requires 
finding a good approximation to the continuum $A_0$ current in 
terms of bilinears made of HISQ light and NRQCD $b$ quarks. 

The most accurate calculation to date is in~\cite{DowdallBdecay} 
where we use a matching through $\alpha_s\Lambda/m_b$, consistent 
with the level of accuracy in our improved NRQCD action:
\begin{equation}
A_0 = (1+z_0\alpha_s)\left[ J_{A_0}^{(0)} 
  + (1+z_1\alpha_s)J_{A_0}^{(1)} + z_2\alpha_s J_{A_0}^{(2)} \right] 
\label{eq:a0match}
\end{equation}
with 
\begin{eqnarray}
J_{A_0}^{(0)} &=& \bar{\Psi}_l \gamma_5 \gamma_0 \Psi_b \\
J_{A_0}^{(1)} &=& \frac{-1}{2m_b} \bar{\Psi}_l \gamma_5 \gamma_0 \gamma \cdot \nabla \Psi_b \nonumber \\
J_{A_0}^{(2)} &=& \frac{-1}{2m_b} \bar{\Psi}_l \gamma \cdot \overleftarrow{\nabla}   \gamma_0 \gamma_5 \gamma_0 \Psi_b . \nonumber 
\end{eqnarray}
The $\mathcal{O}(\alpha_s)$ matching coefficients $z_0$, $z_1$ and $z_2$ 
are given in Table~\ref{tab:zresults}. In~\cite{DowdallBdecay} we 
used $\alpha_s \equiv \alpha_V(n_f=4,2/a)$ to evaluate the matching 
coefficients above. We do that again here and also give 
in Table~\ref{tab:zresults} the value of $\alpha_s$ on each ensemble. 
These are obtained by converting and running down the result 
$\alpha_{\overline{MS}}(n_f=5,M_Z)= 0.1185(6)$~\cite{pdg, McNeile:2010ji}.  

\begin{table}
\caption{\label{tab:zresults} 
Coefficients for the perturbative matching of the 
temporal axial vector and temporal vector currents (eq.~(\ref{eq:a0match})) 
from~\cite{DowdallBdecay}. 
$z_0 = \rho_0-\zeta_{10}$, $z_1 = \rho_1 - z_0$, 
$z_2 = \rho_2$ from~\cite{Monahan:2012dq}.
Column 5 gives the values of $\alpha_s$ used in the 
matching. This is determined in V-scheme with 
4 sea quarks at scale $2/a$. 
}
\begin{ruledtabular}
\begin{tabular}{lllll}
Set & $z_0$ & $z_1$ & $z_2$ & $\alpha_V(2/a)$  \\
\hline \hline
1&0.024(2)  & 0.024(3)  & -1.108(4) & 0.346\\
2&0.022(2)  & 0.024(3)  & -1.083(4) & 0.345 \\
3&0.022(1)  & 0.024(2)  & -1.074(4) & 0.343 \\
4&0.006(2)  & 0.007(3)  & -0.698(4) & 0.311 \\
5&0.001(2)  & 0.007(3)  & -0.690(4) & 0.308 \\
6&0.001(2)  & 0.007(2)  & -0.690(4) & 0.307 \\
7&-0.007(2) & -0.031(4) & -0.325(4) & 0.267 \\
8&-0.007(2) & -0.031(4) & -0.318(4) & 0.266 \\
\end{tabular}
\end{ruledtabular}
\end{table}

To determine the matrix element of $A_0$ and hence $f_B$ we 
calculate the matrix elements of each of the $J_{A_0}^{(i)}$ 
by implementing that operator at the sink of the $B$ meson 
correlation function. For $J_{A_0}^{(0)}$ this is simply 
the local operator used in eq.~(\ref{eq:Bcorr}) (since the 
$\gamma_0$ has no effect on a 2-component NRQCD $b$ quark) 
and $J_{A_0}^{(1)}$ and $J_{A_0}^{(2)}$ are implemented by 
differentiating the appropriate propagator before 
combining into a meson correlator. For a $B$ meson at rest 
the matrix elements of $J_{A_0}^{(1)}$ and $J_{A_0}^{(2)}$ 
are equal since the momenta of the $b$ and $l$ quarks are 
equal and opposite. 

From simultaneous fits, using eq.~(\ref{eq:corrfit}),
 to the $3\times 3$ matrix of $B$ meson correlators 
described above, along with correlators that have $J_{A_0}^{(1)}$ 
inserted at the sink, the amplitude $c(A_0,0)$ that corresponds 
to the annihilation of the ground-state $B$ meson with the $A_0$ current can be 
determined. The amplitude is given by 
\begin{equation}
c(A_0,0) = \frac{\langle 0 | A_0 | B \rangle}{\sqrt{2m_B}} = \frac{f_B\sqrt{m_B}}{\sqrt{2}} .
\label{eq:cdef}
\end{equation}
Results for $\Phi = f_B\sqrt{m_B}$ for $B$ and $B_s$ are given on the sets of 
gluon field ensembles used here in~\cite{DowdallBdecay} and we use 
the same $B$ meson correlation functions here. 

\begin{figure}
\begin{center}
\includegraphics[width=0.9\hsize]{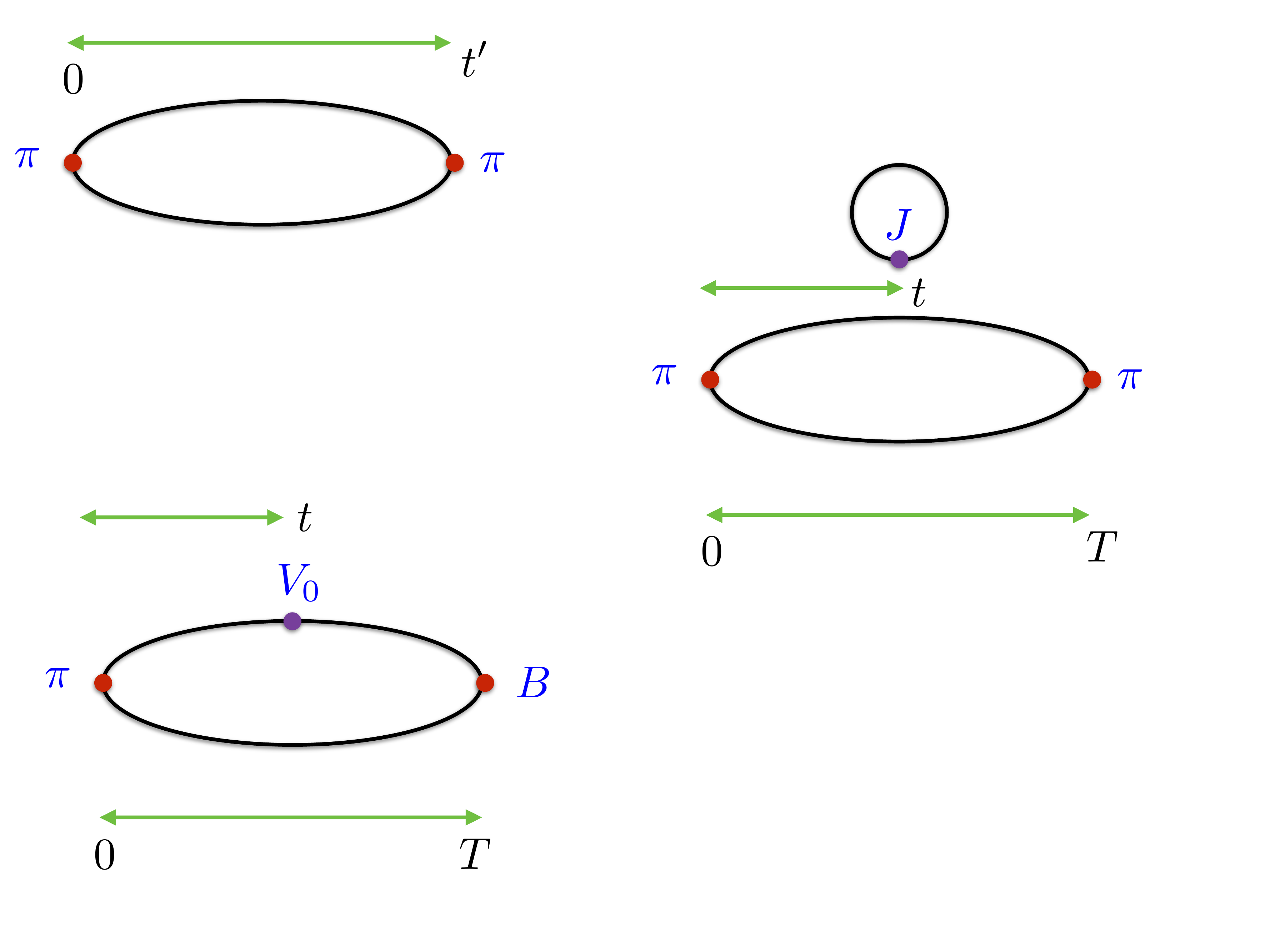}
\end{center}
\caption{ A schematic diagram of the 3-point correlator 
needed for $B \rightarrow \pi \ell \nu$ 
decay at zero recoil. 
}
\label{fig:btopi}
\end{figure}

The new correlation functions that we have calculated here are the 
`3-point' functions for $B \rightarrow \pi \ell \nu$ decay, illustrated in 
Fig.~\ref{fig:btopi}. 
For these a light quark propagator from a random wall source at $t=0$ is 
used, at time-slice $T$ and convoluted with each of the 
$b$-quark smearing functions, as the source of a NRQCD $b$ propagator. 
This propagates backwards in time to time-slice $t$ where it is 
connected via a temporal vector current to another identical light quark propagator, 
joined to the first one at the source time-slice in the 
appropriate way to form a $\pi$ meson. For this to work with staggered 
light quark propagators $\Omega$ matrices must be inserted at $T$ 
and $t$ to reinstate the light quark spin. Both the $B$ meson and 
$\pi$ meson are at rest. We use multiple $T$ values so that we 
can fit the 3-point function both as a function of $t$ and as a function 
of $T$. The $T$ values used on each ensemble are given in Table~\ref{tab:smear}. 

Because of the chiral symmetry of staggered quarks the matching of 
the NRQCD-HISQ temporal 
vector current to continuum QCD takes the same form as the temporal axial vector 
current: 
\begin{equation}
V_0 = (1+z_0\alpha_s)\left[ J_{V_0}^{(0)} 
  + (1+z_1\alpha_s)J_{V_0}^{(1)} + z_2\alpha_s J_{V_0}^{(2)} \right] 
\label{eq:v0match}
\end{equation}
with 
\begin{eqnarray}
\label{eq:v0def}
J_{V_0}^{(0)} &=& \bar{\Psi}_l \gamma_0 \Psi_b \\
J_{V_0}^{(1)} &=& \frac{-1}{2m_b} \bar{\Psi}_l \gamma_0 \gamma \cdot \nabla \Psi_b \nonumber \\
J_{V_0}^{(2)} &=& \frac{-1}{2m_b} \bar{\Psi}_l \gamma \cdot \overleftarrow{\nabla}    \Psi_b . \nonumber 
\end{eqnarray}
We calculate the 3-point function inserting each of the $J_{V_0}^{(i)}$ at the 
heavy-light vertex at $t$. We can then fit the 3-point functions along with 
2-point correlators discussed above to determine the matrix element of 
$V_0$ between $\pi$ and $B$. For $B$ and $\pi$ at rest the 
matrix elements of $J_{V_0}^{(1)}$ and $J_{V_0}^{(2)}$ are equal and 
so we only calculate one of them.  

The 3-point correlation function is fit to the form: 
\begin{eqnarray}
&& C_{3pt}(t,T) = \\
&& \sum_{j,k=0}^{n_{exp}-1} a_{\pi,j}V_{nn}(j,k)c^*_{B,k}fn_{\pi}(E_{\pi,j},t)fn_B(E_{B,k},T-t) \nonumber \\
 &-& \sum_{j,k^{\prime}=0}^{n_{exp}-1} a_{\pi,j}V_{no}(j,k^{\prime})d^*_{B,k^{\prime}}fn_{\pi}(E_{\pi,j},t)fo_B(E^{\prime}_{B,k^{\prime}},T-t) \nonumber
\label{eq:3ptfit}
\end{eqnarray}
where $n$ denotes normal parity states and again oscillating terms appear 
for the $B$ meson from higher mass opposite-parity ($o$) states. Here 
\begin{eqnarray} 
fn_{\pi}(E,t) &=& e^{-Et} + e^{-E(T-t)} ,\\
fn_B(E,t) &=& e^{-Et} , \nonumber \\
fo_B(E,t) &=& (-1)^t fn_B(E,t) .\nonumber 
\label{eq:fndef}
\end{eqnarray}
The amplitudes $a_{\pi,k}$ and energies $E_{\pi, k}$ 
are the same amplitudes and energies as in the $\pi$ 2-point 
fits (eq.~(\ref{eq:pifit})). The amplitudes $c^*_{B,k}$ and $d^*_{B,k^{\prime}}$ 
and energies $E_{B,k}$ and $E^{\prime}_{B,k^{\prime}}$ 
are the same as in the $B$ 2-point fit (eq.~(\ref{eq:corrfit})) for the  
corresponding smearing function for the $B$ meson. 

The fit parameter $V_{nn}(0,0)$ is the result that we need for the ground-state 
$B$ to ground-state $\pi$ matrix element for a current, $J$, inserted 
at $t$. Using the standard relativistic 
normalisation of states: 
\begin{equation}
V^J_{nn}(0,0) = \frac{\langle \pi | J | B \rangle }{2\sqrt{m_{\pi}m_B}}.
\label{eq:vnn}
\end{equation} 
Combining results from $J_{V_0}^{(0)}$, $J_{V_0}^{(1)}$ and $J_{V_0}^{(2)}$ 
as in eq.~(\ref{eq:v0match}) gives an amplitude corresponding 
to the continuum QCD current, $V_0$, which we denote as $V^{V_0}$. 
This is directly related to  
the matrix element of $V_0$ between $\pi$ and $B$ as in eq.~(\ref{eq:vnn}).
Since, at zero recoil
\begin{equation}
\langle \pi | V_0 | B \rangle = f_0(q^2_{max}) (m_B + m_{\pi}) 
\label{eq:matzero}
\end{equation}
then 
\begin{equation}
2\sqrt{m_{\pi}}V^{V_0} = f_0(q^2_{max})\sqrt{m_B}\left(1+\frac{m_{\pi}}{m_B}\right) .
\label{eq:vnnf0}
\end{equation} 

We can therefore directly extract $f_0(q^2_{max})\sqrt{m_B}(1+m_{\pi}/m_B)$ 
from our fit results. Dividing by $\sqrt{2}$ times the amplitude from  
eq.~(\ref{eq:cdef}) gives $f_0(q^2_{max})(1+m_{\pi}/m_B)/f_B$. Note that in 
forming this ratio the overall renormalisation factor of the temporal axial/vector 
current cancels (eqs.~(\ref{eq:a0match}) and~(\ref{eq:v0match})). Hence this 
ratio has significantly lower systematic errors from renormalisation/matching 
than the individual quantities $f_0$ and $f_B$. The division can be done inside 
the fit code and therefore correlations between the fit parameters can be 
taken into account to reduce statistical errors. Because of the inclusion 
of relativistic/radiative corrections to the currents, the ratio is accurate 
through $\alpha_s \Lambda/m_b$ in a power-counting in inverse powers 
of the $b$ quark mass. Multiplication by 
the amplitude and energy combination that 
gives $f_{\pi}$ is also readily done inside the fit to 
give a result for
\begin{equation}
R_{B\pi} = \frac{f_0(q^2_{max})(1+m_{\pi}/m_B)}{[f_B/f_{\pi}]}. 
\label{eq:rdef}
\end{equation}
This is 
the quantity that we will work with, examining its limit as $m_{\pi} \rightarrow 0$ 
where the factor of $(1+m_{\pi}/m_B)$ vanishes and we expect the answer 
1 from the soft-pion relation, eq.~(\ref{eq:softpion}). From this we can also 
obtain the ratio at the physical value of $m_{\pi}$ and thereby the 
value of $f_0(q^2_{max})$ at that point. 

We have also calculated the appropriate 3-point correlation functions for 
the processes $B_s \rightarrow K \ell \nu$ and $B_s \rightarrow \eta_s \ell \nu$ 
and combined these with the appropriate 2-point functions 
from~\cite{DowdallBdecay, Dowdallfkpi} to obtain an analogous ratio 
for $f_0(q^2_{max})$ to that above ($R_{B_sK}$ and $R_{B_s\eta_s}$ respectively). 
In these cases we can also extract 
a result for $f_0(q^2_{max})$ for physical quark masses that is accurate 
through $\alpha_s\Lambda/m_b$. 

\section{Results}
\label{sec:results}

\subsection{$B \rightarrow \pi$}
\label{subsec:bpi}
As discussed in Section~\ref{sec:latt}, we fit our results for 3-point 
functions for $B \rightarrow \pi$ 
and 2-point functions for $B$ and $\pi$
simultaneously to the forms given in eqs.~(\ref{eq:pifit}), (\ref{eq:corrfit})
and (\ref{eq:3ptfit}). We use a constrained fitting technique~\cite{gplbayes} 
so that we can include uncertainties in our fitted results for the ground-state 
coming from the presence of excited states in the correlation function. 
The prior value taken on the difference in mass between adjacent states (both 
in normal and oscillating channels) 
is 600(300) MeV and (for the $B$) on the difference between the ground-state 
and the first oscillating state is 400(200) MeV. The prior taken on all 
2-point amplitudes is 0.0(1.0) and for 3-point amplitudes 0.0(5.0). 
For our normalisation of the raw correlators these 
widths correspond to 3--5 times the ground-state value and so provide a loose 
constraint. 

\begin{figure}
\begin{center}
\includegraphics[width=\hsize]{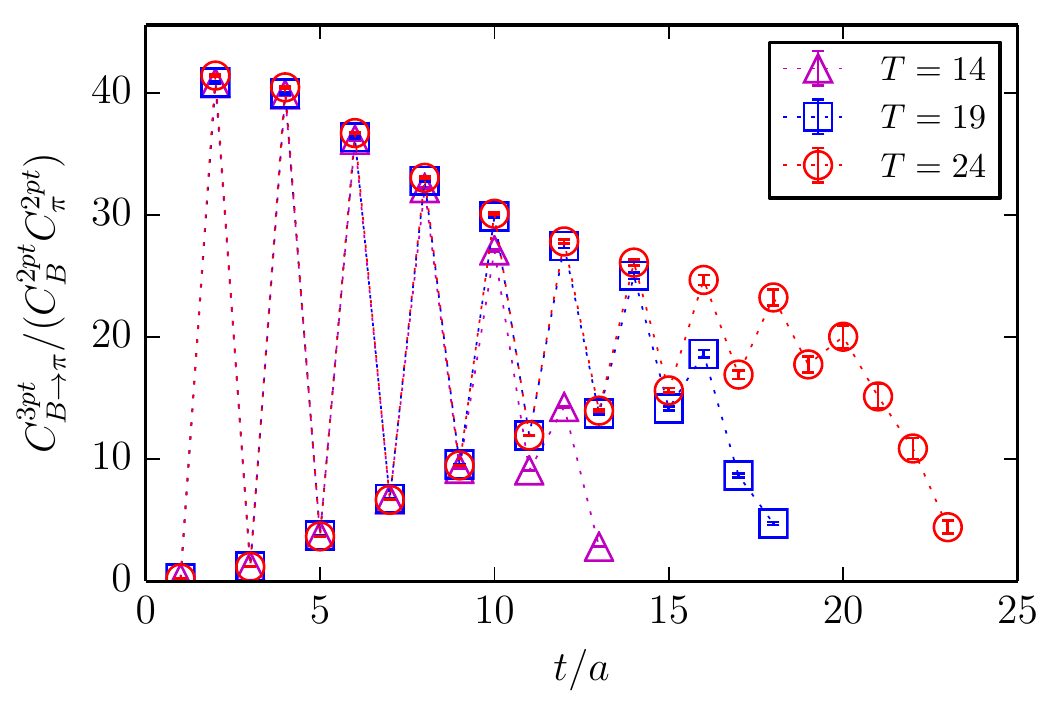}
\end{center}
\caption{ The ratio of the 3-point correlator for $B \rightarrow \pi$ decay 
to the product of 2-point correlators for $B$ and $\pi$ as a function of 
lattice time. The $B$ is at 
$t=0$ and the $\pi$ at $T$ here, since this illustrates more clearly the 
convergence of results for different $T$. 
Results are for coarse set 4, statistical errors only.
}
\label{fig:ratplot}
\end{figure}

Figure~\ref{fig:ratplot} illustrates the quality of our results by showing  
a plot of the ratio of 3-point to 2-point correlators for 
$B \rightarrow \pi$ decay on coarse set 4. Statistical errors are small
and, although oscillating terms are strong on the $B$ side of 
the 3-point correlator, it is clear that results at different $T$ values 
are consistent so that we are able to isolate ground-state 
amplitudes.  

\begin{table}
\begin{tabular}{lllll}
\hline
\hline
Set &  $V^{(0)}_{nn}(0,0)$ & $V^{(1)}_{nn}(0,0)$ & $V^{V_0}$ & $g^{B\pi}_0(q^2_{\mathrm max})$ \\
\hline
1 & 2.072(33) & -0.115(8)& 2.017(33) & 1.961(32)\\
2 & 2.581(40) & -0.159(5)& 2.499(40) & 2.037(33) \\
3 & 3.637(49) & -0.247(21)& 3.505(61) & 2.236(39) \\
\hline
4 & 2.110(32) & -0.129(3)& 2.012(31) & 1.761(27) \\
5 & 2.675(27) & -0.182(3)& 2.532(27) & 1.855(20) \\
6 & 3.837(50) & -0.289(5)& 3.609(50) & 2.061(28) \\
\hline
7 & 2.154(17) & -0.156(2) & 2.010(17) & 1.508(13) \\
8 & 3.829(91) & -0.334(10) & 3.520(85) & 1.683(41) \\
\hline
\hline
\end{tabular}
\caption{Columns 2 and 3 give ground-state parameters $V_{nn}(0,0)$ 
(in lattice units) from our combined 2-point and 3-point 
fit for $B \rightarrow \pi$ decay. 
$V^{(0)}_{nn}$ corresponds 
to current $J_{V_0}^{(0)}$ and $V^{(1)}_{nn}$ to $J_{V_0}^{(1)}$ 
(see eq.~(\ref{eq:v0def})). $J_{V_0}^{(2)}$ has equal amplitude 
to $J_{V_0}^{(1)}$ at zero recoil and so is not given 
separately. In column 4 results are combined as 
in eq.~(\ref{eq:v0match}) into an equivalent parameter for 
the full QCD current, $V_0$. Column 5 presents the results as 
the combination $g^{B\pi}_0 \equiv f_0\sqrt{am_B}(1+m_{\pi}/m_B)$ in 
lattice units. }
\label{tab:bpiresV}
\end{table}

\begin{table}
\centerline{
\begin{tabular}{lcc}
\hline
\hline
Set & $\frac{f^{(0)}_{0}(q^2_{\mathrm{max}})(1+m_{\pi}/m_B)}{(f^{(0)}_B/f_\pi)}$ & $R_{B\pi} \equiv \frac{f_{0}(q^2_{\mathrm{max}})(1+m_{\pi}/m_B)}{(f_B/f_\pi)}$ \\
\hline
1 & 0.696(11) & 0.708(11) \\
2 & 0.726(12) & 0.735(13) \\
3 & 0.777(12) & 0.783(13) \\
\hline
4 & 0.712(12) & 0.728(13) \\
5 & 0.729(8) & 0.740(9) \\
6 & 0.806(11) & 0.814(12) \\
\hline
7 & 0.703(6) & 0.721(7) \\
8 & 0.777(24) & 0.786(25) \\
\hline
\hline
\end{tabular}
}
\caption{Results for the ratio of $f_0(q^2_{\mathrm max})$ for 
$B \rightarrow \pi$ decay to 
$f_B/f_{\pi}$ in the combination that appears naturally from 
our fits: $R_{B\pi} \equiv f_0(q^2_{\mathrm max})(1+m_{\pi}/m_B)/[f_B/f_{\pi}]$. 
Column 3 gives the full result and column 2 the result from 
using the leading order NRQCD-HISQ current only, both in 
$f_0$ and in $f_B$. }
\label{tab:bpifrat}
\end{table}

The results from our 2-point fits have been detailed in~\cite{DowdallBdecay} 
and~\cite{Dowdallfkpi} and we obtain results in good agreement with those 
values here. Table~\ref{tab:bpiresV} gives values for the 
ground-state $B$ to ground-state $\pi$ 3-point amplitude $V_{nn}(0,0)$ from 
eq.~(\ref{eq:3ptfit}) for the case where the current inserted is $J_{V_0}^{(0)}$ 
and $J_{V_0}^{(1)}$ (these two cases are fit simultaneously). 
We see that the raw matrix element for the sub-leading current is 
5--8\% of the leading current. We also give results for the 
combination $V^{V_0}$ from eq.~(\ref{eq:v0match}) that corresponds to the full 
QCD current (to the order to which we are working), $V_0$.  
The final column of Table~\ref{tab:bpiresV} uses eq.~(\ref{eq:vnnf0}) 
to determine the combination $f_0(q^2_{\mathrm max})\sqrt{am_B}(1+m_{\pi}/m_B)$. 
We see significant dependence on the mass of the 
$\pi$ meson in these results, a feature that was missing in 
earlier calculations that could not reproduce the soft pion 
theorem relationship for $f_0(q^2_{\mathrm max})$~\cite{Hashimoto:1999bk}. 

\begin{figure}
\begin{center}
\includegraphics[width=\hsize]{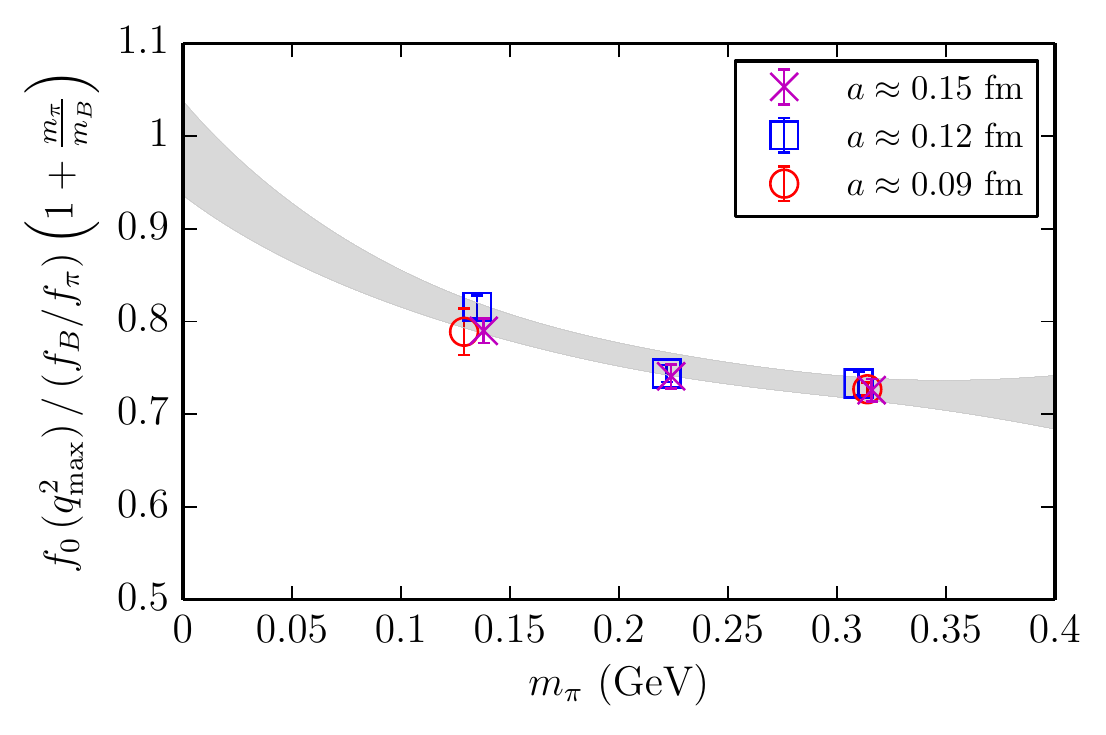}
\end{center}
\caption{Results for the ratio $R_{B\pi}$ (eq.~(\ref{eq:rdef})) 
of form factor $f_0(q^2_{max})$ (multiplied by $(1+m_{\pi}/m_B)$) 
to $f_B/f_{\pi}$, plotted as a function of $m_{\pi}$.  
Results from very coarse ensembles are shown as crosses, coarse as open 
squares and fine as open circles. These include a small correction 
for finite volume effects, as discussed in the text.  
The grey band shows the fit described in the text, agreeing 
with value 1 at $m_{\pi}=0$. 
}
\label{fig:fplot}
\end{figure}

As discussed in Section~\ref{sec:latt}, 
we can also evaluate, directly from our fits, the
combination $f_0(q^2_{max})(1+m_{\pi}/m_B)/[f_B/f_{\pi}]$ 
by dividing our 3-point amplitude by appropriate 
2-point amplitudes which are correlated within the fit. 
This dimensionless ratio, in which the overall renormalisation 
factor for the lattice currents cancels, is tabulated in 
Table~\ref{tab:bpifrat}. There we give the result both 
the full calculation, using currents $V_0$ in the 3-point amplitude and
$A_0$ in $f_B$, and for the calculation using just 
the leading order current $J^{(0)}$ in both cases.  
The difference between the two is small, around 2\%, 
because the impact of the sub-leading currents largely 
cancels between $f_0(q^2_{\mathrm max})$ and $f_B$. 
Our statistical/fitting uncertainty is about at the 
same level.  The results for the full ratio 
are plotted in Figure~\ref{fig:fplot} as a function 
of $m_{\pi}$. Again, dependence on $m_{\pi}$ is clear, 
but no dependence on the lattice spacing is seen since
discretisation effects evident for $f_B$ in~\cite{DowdallBdecay} largely 
cancel in the ratio.  

In order to test the soft pion theorem we need to 
fit our results as a function of $m_{\pi}$ to 
extrapolate to $m_{\pi}=0$. Having results for a 
range of small $m_{\pi}$ values allows us to do this. 

One key element of $m_{\pi}$ dependence is purely 
kinematic~\cite{Bowler:1999xn}: the fact that $q^2_{\mathrm max}$ depends on 
$m_{\pi}$ through the formula 
\begin{equation}
q^2_{\mathrm max} = (m_B-m_{\pi})^2 .
\end{equation}
This will mean that, if results for different $m_{\pi}$ 
fall on similar $f_0$ curves as a function of $q^2$, 
there will be a significant linear term in $m_{\pi}$ 
coming from the shift in $q^2_{\mathrm max}$ as 
$m_{\pi}$ is reduced.  A slope in $m_{\pi}$ of 
order $-2m_B df_0/dq^2|_{q^2=q^2_{\mathrm max}}$ 
might then be expected. An estimate for this 
slope can be derived from a simple pole form for $f_0(q^2)$, 
behaving as $f_0(0)/[1-q^2/m^2_{B^*_0}]$, where 
the $B^*_0$ is the lowest scalar state in the $B$ meson 
spectrum. This would give 
a slope in $m_{\pi}$ of $f_0(q^2_{max})/\Lambda$ where 
$\Lambda$ is the difference in mass between the 
$B^*_0$ and the $B$. Taking $\Lambda$ of order 500 MeV 
results in a slope 
of 2 $\mathrm{GeV}^{-1}$. The term in $m_{\pi}^2$ coming 
from the kinematics above would have a much smaller slope, 
of $\mathcal{O}(1/(2\Lambda m_B))$, 
since there is no enhancement by $m_B$.    
This slope is then smaller than that for powers of $m_{\pi}$ 
coming from chiral perturbation theory and so can be 
absorbed into those terms in our fit. 

From low-order chiral perturbation theory we expect 
simple powers of $m_{\pi}^2$ as well as chiral logarithms 
of the form 
$m_{\pi}^2/(4\pi f_{\pi})^2 \times \log(m^2_{\pi}/\Lambda^2_{\chi})$ 
(where we take $f_{\pi} =$ 130 MeV). 
The appropriate 
staggered quark chiral perturbation theory 
for these quantities 
is given in~\cite{Aubin:2003uc, Aubin:2005aq, Aubin:2007mc}, 
see also~\cite{Becirevic:2003ad}. The way in which the 
masses of different tastes of $\pi$ meson appear in the 
chiral logarithm terms and associated `hairpins' is 
discussed in Appendix~\ref{appendix:schipt}.  
It turns out that the staggered chiral perturbation theory 
for $f_{\pi}$ and $f_B$ is very like the continuum 
form, because discretisation effects from the 
masses of $\pi$ mesons of different taste almost 
entirely cancel out. 
Terms of this kind in fact cancel between 
$f_0$ and $f_B/f_{\pi}$. This includes in 
particular all the chiral 
logarithms with coefficient $g_{B^*B\pi}$.  

The remaining chiral logarithms that do not cancel 
in the ratio of $f_0(q^2_{max})$ and $f_B/f_{\pi}$ 
take the form of an average over tastes, $t$, of 
$m_{\pi_G}^2/(4\pi f_{\pi})^2 \times \log(m^2_{\pi,t}/\Lambda^2_{\chi})$. 
Here $m_{\pi_G}$ is the mass of the Goldstone $\pi$ 
meson in the final state, and $m_{\pi,t}$ is the mass of 
a $\pi$ of taste $t$. 
$m_{\pi_G} \equiv m_{\pi}$ is the expansion parameter for 
the chiral perturbation theory since its square 
is proportional to the light quark mass.  
The masses of the other taste $\pi$ mesons, $m_{\pi,t}$, 
could contribute discretisation errors to the chiral 
perturbation theory, when compared to the continuum 
chiral logarithms, since they differ by 
$\alpha_s^2a^2$ from $m_{\pi_G}$. These discretisation 
errors would be relatively benign, since the chiral 
logarithm above
vanishes as $m_{u/d} \rightarrow 0$ 
even at non-zero lattice spacing. 
In fact, as shown in Appendix~\ref{appendix:schipt}, 
hairpin diagrams also cancel most of these discretisation 
effects giving a dependence on $m_{\pi}$ which is almost 
identical to that of continuum chiral perturbation theory. 
We therefore use continuum chiral perturbation theory 
for our extrapolation to $m_{\pi} = 0$ 
but allow for $m_{\pi}$-dependent discretisation errors to 
include remaining staggered taste-changing effects. 

Combining results for $f_B$, $f_{\pi}$ 
and $f_0$ gives a coefficient for the chiral logarithm 
above
in $f_0f_{\pi}/f_B$ of -4. 
A value of -4 is a substantial coefficient, double that 
in $f_{\pi}$ for example, and so it might be expected that we 
need to pay attention to finite-volume effects. 
We studied these 
for $f_{\pi}$ on these ensembles in~\cite{Dowdallfkpi} and found 
them to be well below 1\%, except on set 1 which has the coarsest 
lattice spacing and is furthest from the physical point and so has 
very little impact on any fit. Finite-volume corrections were included 
in that study because they were significant at the level of the 
statistical errors possible there. Here statistical errors 
are much larger and we find finite volume effects are not significant. 
We include double the 
relative finite volume effect seen in $f_{\pi}$ for the 
ratio $f_0f_{\pi}/f_B$ in the values shown in Figure~\ref{fig:fplot}. 
This has negligible impact on the final 
result at physical $m_{\pi}$ and less than 1\% effect on 
the value at $m_{\pi} = 0$.    

\begin{table}
\begin{tabular}{l | ccc}
\hline
\hline
 & $R_{B\pi}$ & $R_{B_sK}$ & $R_{B_s\eta_s}$ \\
\hline
stats/fitting & $1.4$ & $0.7$ & $0.6$ \\
$a$-dependence & $1.3$ & $0.6$ & $0.4$ \\
NRQCD systematics & $0.03$ & $0.03$ & $0.04$ \\
$m_{\pi}$ dependence & $0.5$ & $0.2$ & $0.1$ \\
$s$ quark tuning & - & $0.03$ & $0.01$ \\
\hline
Total (\%) & $1.9$ & $1.0$ & $0.7$ \\
\hline
\hline
\end{tabular}
\caption{Errors budgets for the ratios $R_{B\pi}$, $R_{B_sK}$ and 
$R_{B_s\eta_s}$ (defined in eqs~(\ref{eq:rdef}),~(\ref{eq:rdefK}) 
and~(\ref{eq:rdefetas})) evaluated at the physical value of $m_{\pi}$. 
Uncertainties are given as a percentage of the final answer. }
\label{tab:errs}
\end{table}

We fit the results for the ratio $R_{B\pi}$ as a function of lattice spacing and 
$\pi$ meson mass to the following form: 
\begin{eqnarray}
\label{eq:fitform}
R_{B\pi}(a,m_{\pi}) &=& R_{B\pi}(\mathrm{phys}, m_{\pi}=0)\times\left[ 1  \right. \\ 
&+& \sum_{j=1,3} c_j(a\Lambda)^{2j} + d \delta_{a^2,m^2_{\pi}} \nonumber \\
&+& \left(\frac{\Lambda}{m_b}\right)^2(a\Lambda)^2[e_{1}\delta x_m + e_{2}(\delta x_m)^2] \nonumber \\ 
&+& \left. \sum_{k=1,4} f_k \left(\frac{m_{\pi}}{\Lambda_{\chi}}\right)^k -g\left(\frac{m_{\pi}}{4\pi f_{\pi}}\right)^2\log{\frac{m_{\pi}^2}{\Lambda_{\chi}^2}} \right] \nonumber 
\end{eqnarray}
with $\Lambda_{\chi}$ = 1 GeV. 
Here the $c_j$ coefficients provide for regular discretisation errors which 
appear for our actions as even powers of $a$ only. We take $\Lambda$ = 400 MeV. 
The coefficient $d$ provides for light quark mass-dependent discretisation errors,
as discussed above; more information on these is given below.
The terms with coefficients $e_1$ and $e_2$ allow for discretisation errors from 
higher-order terms in the NRQCD action with coefficients that might depend on 
$am_b$. $\delta x_m = (am_b-2.7)/1.5$, so it varies from -0.5 to 0.5 
across the range used here for $am_b$. We take $\Lambda/m_b=0.1$. 
The $f_k$ coefficients provide for $m_{\pi}$ dependence expected from 
kinematic effects from the dependence of $q^2_{\mathrm max}$ on 
$m_{\pi}$ (which includes in particular a linear term in $m_{\pi}$ 
as discussed above). Any dependence on (even) powers of $m_{\pi}$ from 
chiral perturbation theory will be subsumed into this dependence. 
The final term is a chiral logarithm, for which 
we allow coefficient $g$. As discussed above, we use the continuum 
form for the chiral logarithm and allow for $m_{\pi}$-dependent 
discretisation effects from staggered chiral perturbation theory 
with the $\delta_{a^2,m_{\pi}^2}$ term. Our fits are insensitive to 
the form that this term takes. We have tried, for example, a form 
based on discretisation effects from averaging over $\pi$ meson tastes in 
a logarithm: 
\begin{equation}
\delta_{a^2,m_{\pi}^2} = \frac{m_{\pi}^2}{(4\pi f_{\pi})^2}\log(1+2\delta_t/m_{\pi}^2) .
\end{equation} 
$\delta_t$ is one unit of taste-splitting (see Appendix~\ref{appendix:schipt}). 
For our final results we in fact use the simpler 
$\delta_{a^2,m_{\pi}^2}=a^2m^2_{\pi}$. In neither 
case does the fit return a significant coefficient for this term. 

We take a prior of 0.0(1.0) on almost all coefficients in eq.~(\ref{eq:fitform}). 
The exceptions are: a prior of 1.0(5) on $R_{B\pi}(\mathrm{phys}, m_{\pi}=0)$; a prior 
of $0.0(3)$ on $c_1$, since tree-level $a^2$ errors are missing from our action 
so that the leading term is $\alpha_sa^2$; a prior of 0.0(5.0) on $f_1$ since 
this linear term in $m_{\pi}$ is expected (from the arguments above) 
to have a coefficient around 2 $\mathrm{GeV}^{-1}$
from its origin in the $m_{\pi}$ dependence of $q^2_{\mathrm max}$ and 
a prior of 4.0(1.0) on $g$, the coefficient of the chiral logarithm, which 
is known up to chiral corrections.     

The chiral fit has a $\chi^2/{\mathrm{dof}}$ of 0.7 for 8 degrees of freedom 
and gives the result $R_{B\pi}(\mathrm{phys},m_{\pi}=0)$ = 0.987(51), 
i.e. 1 in agreement with the soft pion relation of eq.~(\ref{eq:softpion}) 
with an uncertainty of 5\%. Most of the parameters in the fit are not 
well determined by the data, except for the linear term in $m_{\pi}$ which 
has coefficient -2.4(5), in agreement with our expectation. 
Missing out the chiral logarithm changes things slightly, giving a result 
2$\sigma$ below 1 at $m_{\pi}$=0 of 0.92(4). 
Leaving the chiral logarithm unconstrained, i.e. giving $g$ a prior 
of 0.0(10.0), results in a fitted value for $g$ of 2(8), i.e. it is 
not well determined by the results (but also not inconsistent with its 
expected value).  

From the fit we can extract the result at the physical value of 
the $\pi$ meson mass corresponding to equal mass $u$ and $d$ quarks 
in the absence of electromagnetism. This is the experimental mass 
of the $\pi^0$, 135 MeV. There we find 
\begin{equation}
R_{B\pi}(\mathrm{phys},m_{\pi}=m_{\pi^0}) = 0.805(16). 
\end{equation}
This is very insensitive to any of the details of the fit because we have 
results at the physical $\pi$ mass.  The error budget for 
$R_{B\pi}(\mathrm{phys},m_{\pi}=m_{\pi^0})$ is given in Table~\ref{tab:errs}. 
Using our value of $f_B$ = 0.186(4) GeV~\cite{DowdallBdecay} (for the $m_u=m_d=m_l$ case) 
and the experimental value for $f_{\pi^+}$ = 130.4 MeV, along with 
$(1+m_{\pi^0}/m_B)$ = 1.0256 gives 
\begin{equation}
f_0(q^2_{max})|_{B\rightarrow \pi} = 1.120(22)(24) .
\label{eq:f0res}
\end{equation} 
The first uncertainty comes from the fit to the ratio $R_{B\pi}$ and 
includes statistical and fitting errors. The second uncertainty comes 
from our value for $f_B$ and includes uncertainties from missing 
higher order terms in the current matching. We expect these to be very 
similar for $f_0$, since we have seen this to be the case for 
currents we have included, so we do not include any additional systematic error
specific to $f_0$. We find finite-volume effects, as discussed 
above, to give negligible uncertainty.   

Taking instead our value of $f_{B^+}$ of 0.184(4) GeV~\cite{DowdallBdecay} 
gives  
\begin{equation}
f_0(q^2_{max})|_{B^+\rightarrow \pi^0} = 1.108(22)(24) .
\label{eq:f0res2}
\end{equation} 

\subsection{$B_s \rightarrow \eta_s$}
\label{subsec:bsetas}

\begin{table}
\begin{tabular}{llllll}
\hline
\hline
Set &  $V^{(0)}_{nn}(0,0)$ & $V^{(1)}_{nn}(0,0)$ & $V^{V_0}$ & $g^{B_s\eta_s}_0(q^2_{\mathrm max})$ & $R_{B_s\eta_s}$ \\
\hline
1 & 1.329(14) & -0.042(2)& 1.314(14) & 1.886(21) & 0.740(8)\\
2 & 1.323(10) & -0.042(1)& 1.306(10) & 1.867(15)  & 0.743(6)\\
\hline
4 & 1.325(7) & -0.050(0)& 1.267(7) & 1.648(9)  & 0.733(4)\\
6 & 1.322(2) & -0.051(0)& 1.282(2) & 1.652(3)  & 0.740(1)\\
\hline
8 & 1.347(6) & -0.066(0) & 1.284(6) & 1.418(6)  & 0.742(3)\\
\hline
\hline
\end{tabular}
\caption{Columns 2 and 3 give ground-state parameters in lattice units 
$V_{nn}(0,0)$ 
from our combined 2-point and 3-point fit for 
$B_s \rightarrow \eta_s \ell \nu$. 
$V^{(0)}_{nn}$ corresponds 
to current $J_{V_0}^{(0)}$ and $V^{(1)}_{nn}$ to $J_{V_0}^{(1)}$ 
(see eq.~(\ref{eq:v0def})). $J_{V_0}^{(2)}$ has equal amplitude 
to $J_{V_0}^{(1)}$ at zero recoil and so is not given 
separately. In column 4 results are combined as 
in eq.~(\ref{eq:v0match}) into an equivalent parameter for 
the full QCD current, $V_0$. Column 5 presents the results as 
the combination 
$g^{B_s\eta_s}_0 \equiv f_0\sqrt{am_{B_s}}(1+m_{\eta_s}/m_{B_s})$ in 
lattice units and Column 6 gives the ratio $R_{B_s\eta_s}$ defined 
in eq.~(\ref{eq:rdefetas}). }
\label{tab:betares}
\end{table}

\begin{figure}
\begin{center}
\includegraphics[width=\hsize]{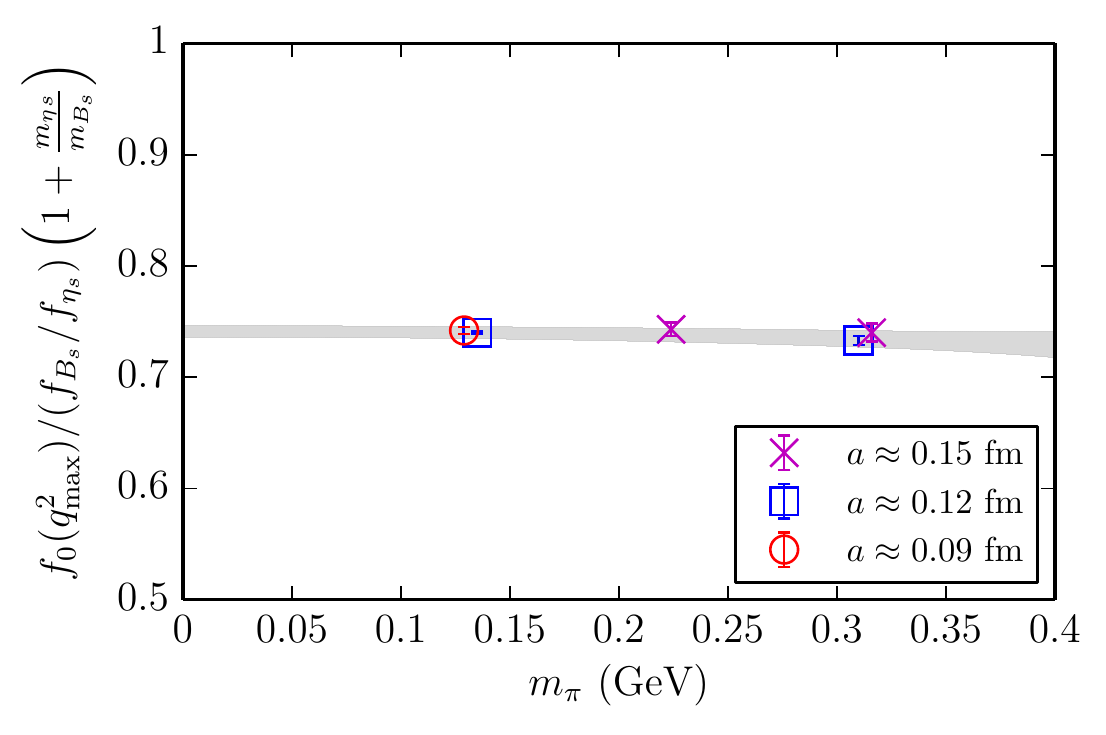}
\includegraphics[width=\hsize]{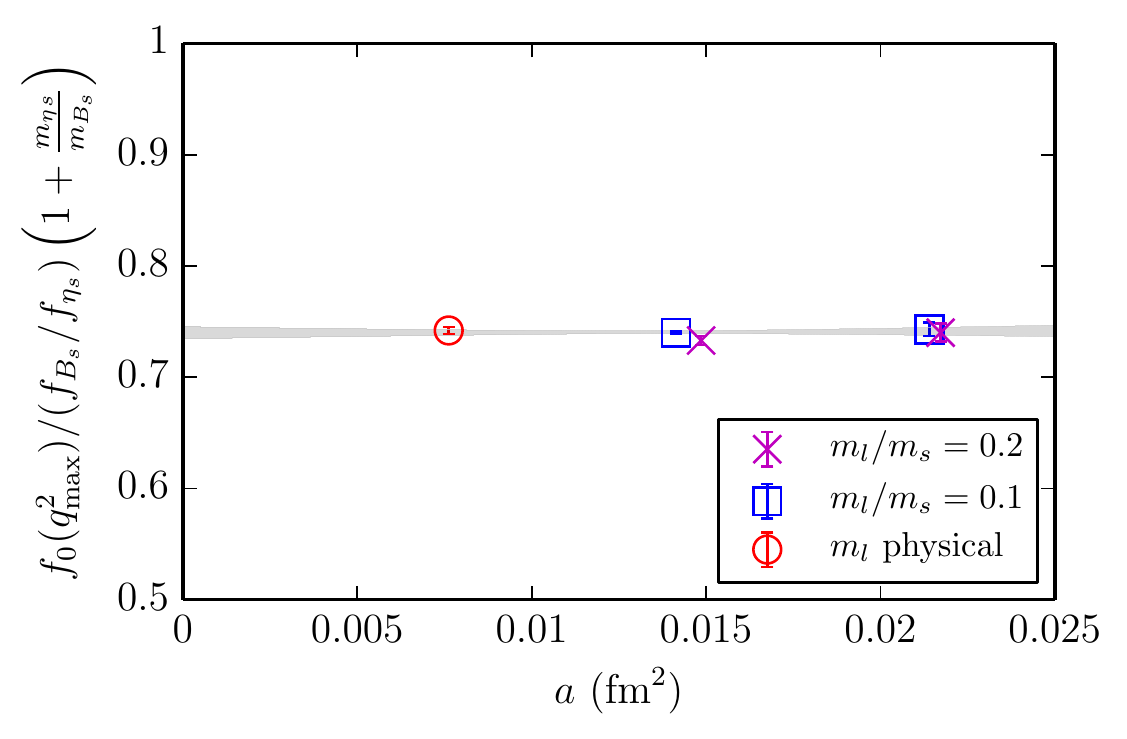}
\end{center}
\caption{Results for the ratio $R_{B_s\eta_s}$ of eq.~(\ref{eq:rdefetas})
plotted against $m_{\pi}$ in the top plot and against the 
lattice spacing in the lower plot (see Table~\ref{tab:betares}). 
The grey band gives the results of a fit described in the text. 
}
\label{fig:etasfplot}
\end{figure}

The decay mode $B_s \rightarrow \eta_s$, in which we replace the 
$l$ quark in $B \rightarrow \pi$ with an $s$ quark is a useful 
calibration point in lattice QCD. The $\eta_s$ meson is a pseudoscalar 
meson with valence quark content $s\overline{s}$ 
but including the quark-line-connected 
correlator only so that it cannot mix with other flavour-singlet mesons. 
It is thus not a physical particle that appears in experiment. However 
its properties, mass and decay constant, have been well-studied in 
lattice QCD~\cite{Davies:2009tsa, Dowdallfkpi}. 
 
The calculation for $B_s \rightarrow \eta_s$ proceeds exactly 
as described in Section~\ref{sec:latt} for $B \rightarrow \pi$. 
The valence $s$ quarks have the masses in lattice units given 
in Table~\ref{tab:mbc}. 2-point correlators for the 
$B_s$ and $\eta_s$ were calculated in~\cite{DowdallBdecay, Dowdallfkpi}. 
In Table~\ref{tab:betares} we give the results from this calculation for the 
3-point amplitude for $B_s \rightarrow \eta_s$ decay via a temporal 
vector current with $B_s$ and $\eta_s$ at rest. 
We also give the result derived directly from our fits for the ratio 
\begin{equation}
R_{B_s\eta_s} = \frac{f_0(q^2_{max})(1+m_{\eta_s}/m_{B_s})}{[f_{B_s}/f_{\eta_s}]}. 
\label{eq:rdefetas}
\end{equation}
in which the overall renormalisation of currents cancels. 
The ratio $R_{B_s\eta_s}$ is plotted in Figure~\ref{fig:etasfplot} as a function 
of the mass of the $\pi$ meson made from the sea $l$ quarks and 
as a function of the lattice spacing, $a$. 
We have results only for a subset of the ensembles used for 
$B \rightarrow \pi$, but we see no dependence of 
$R_{B_s\eta_s}$ on either $m_{\pi}$ or on $a$. 

To fit the results for $B_s \rightarrow \eta_s$ decay and 
derive a physical result for $f_0(q^2_{\mathrm max})$ we 
use a similar fit form to that of eq.~(\ref{eq:fitform}).
We drop the chiral logarithm term as well as odd powers 
of $m_{\pi}$ from kinematic effects, since these are no longer 
relevant. We add a term allowing for mistuning of the 
$s$ quark mass as $h(m_{\eta_s}^2 - [0.6885\,\mathrm{GeV}]^2)$ with prior 
on h of 0.0(2). Here 0.6885 GeV is the 
the $\eta_s$ mass in the continuum and chiral limits 
determined from lattice QCD~\cite{Dowdallfkpi}. 

Our fit has a $\chi^2/{\mathrm{dof}}$ of 0.35 for 5 degrees of 
freedom and gives a result at physical $m_{\pi}$ for $R_{B_s\eta_s}$ of 
0.740(5). 
For this quantity the value at $m_{\pi}=0$ has no significance. 
The physical result for $R_{B_s\eta_s}$ is very 
insensitive to any details of the fit 
since we have results at the physical $\pi$ mass and there is 
almost no dependence on $m_{\pi}$ and $a$. The error budget is given 
in Table~\ref{tab:errs}; the uncertainty is dominated by 
statistics. The value of 
$R_{B_s\eta_s}$ can be converted into a value for 
$f_0(q^2_{max})$ using $f_{B_s}$ = 0.224(5) GeV~\cite{DowdallBdecay}
and $f_{\eta_s}$ = 0.1811(6) GeV and 
$(1+m_{\eta_s}/m_{B_s})$ = 1.1283~\cite{Dowdallfkpi, pdg}.  
We find 
\begin{equation}
\left. f_0(q^2_{max})\right|_{B_s\rightarrow \eta_s} = 0.811(5)(16)
\label{eq:f0resetas}
\end{equation} 
where the first error is from the ratio 
and the second from $f_{B_s}$ and $f_{\eta_s}$. 
This is a substantially smaller value than that for $B \rightarrow \pi$ 
and it is also more precise, reflecting smaller statistical uncertainties 
in the raw data and the absence of any dependence on $m_{\pi}$ or $a$. 

\subsection{$B_s \rightarrow K$}
\label{subsec:BsK}

\begin{table}
\begin{tabular}{llllll}
\hline
\hline
Set &  $V^{(0)}_{nn}(0,0)$ & $V^{(1)}_{nn}(0,0)$ & $V^{V_0}$ & $g^{B_sK}_0(q^2_{\mathrm max})$ & $R_{B_sK}$ \\
\hline
1 & 1.490(8) & -0.060(1)& 1.465(8) & 1.854(10) & 0.656(4) \\
2 & 1.510(16) & -0.064(4)& 1.481(16) & 1.825(20)  & 0.639(7) \\
\hline
4 & 1.496(9) & -0.069(1)& 1.445(9) & 1.654(10)  & 0.662(4) \\
6 & 1.539(8) & -0.075(1)& 1.480(8) & 1.617(9)  & 0.623(3) \\
\hline
8 & 1.571(9) & -0.094(2) & 1.483(9) & 1.389(9)  & 0.623(4) \\
\hline
\hline
\end{tabular}
\caption{Columns 2 and 3 give ground-state parameters in lattice units 
$V_{nn}(0,0)$ 
from our combined 2-point and 3-point fit for 
$B_s \rightarrow K \ell \nu$. 
$V^{(0)}_{nn}$ corresponds 
to current $J_{V_0}^{(0)}$ and $V^{(1)}_{nn}$ to $J_{V_0}^{(1)}$ 
(see eq.~(\ref{eq:v0def})). $J_{V_0}^{(2)}$ has equal amplitude 
to $J_{V_0}^{(1)}$ at zero recoil and so is not given 
separately. In column 4 results are combined as 
in eq.~(\ref{eq:v0match}) into an equivalent parameter for 
the full QCD current, $V_0$. Column 5 presents the results as 
the combination $g^{B_sK}_0 \equiv f_0\sqrt{am_{B_s}}(1+m_{K}/m_{B_s})$ in 
lattice units and Column 6 gives the ratio $R_{B_sK}$ defined 
in eq.~(\ref{eq:rdefK}). }
\label{tab:bskres}
\end{table}

\begin{figure}
\begin{center}
\includegraphics[width=\hsize]{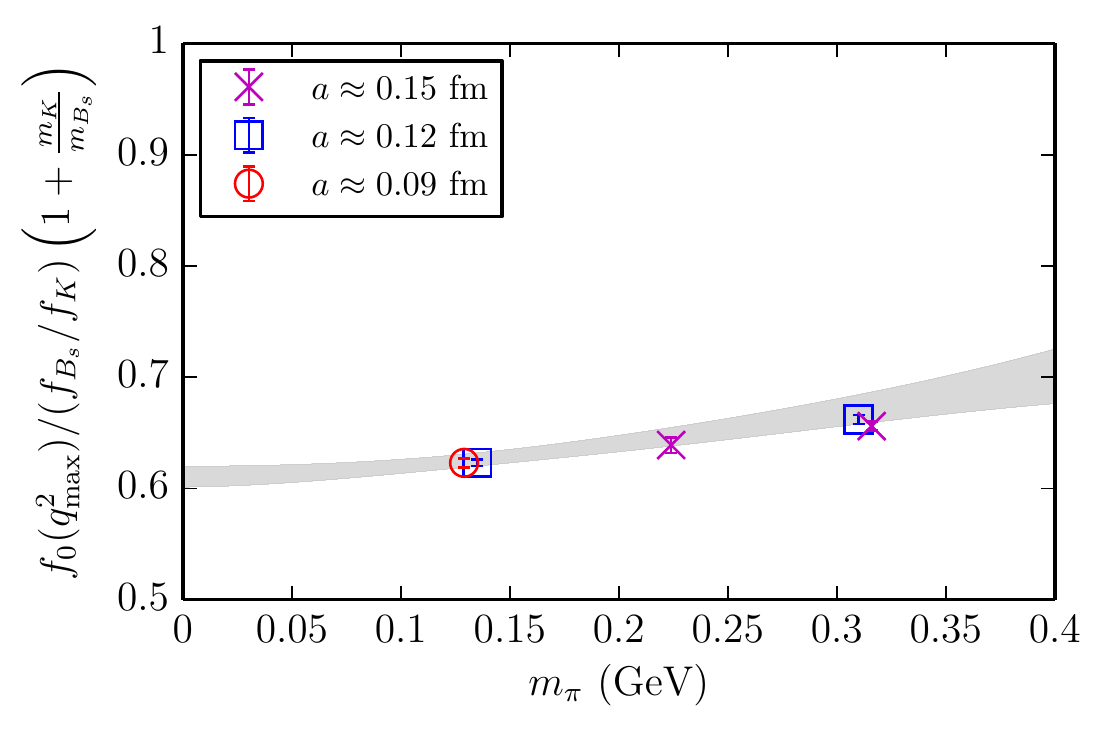}
\end{center}
\caption{The ratio $R_{B_sK}$ of eq.~(\ref{eq:rdefK})
plotted against $m_{\pi}$. 
See Table~\ref{tab:bskres} for results. 
The grey band gives the results of a fit described in the text. 
}
\label{fig:Kfplot}
\end{figure}

The decay mode $B_s \rightarrow K \ell \nu$ is a physical process 
in which a $b$ quark undergoes a weak transition to a $u$ quark 
inside a $B_s$ meson. This process can then be used in a similar way 
to $B \rightarrow \pi \ell \nu$ to determine $V_{ub}$. 
Lattice QCD calculations for the form factors for this process are 
now starting to appear~\cite{Bouchard:2014ypa, RBCbtopi2015}. 
Here we again give results for 
the form factor $f_0$ at zero recoil with $u/d$ quark masses going 
down to physical values. 
We again study the ratio determined directly from our fits:
\begin{equation}
R_{B_sK} = \frac{f_0(q^2_{max})(1+m_{K}/m_{B_s})}{[f_{B_s}/f_{K}]}. 
\label{eq:rdefK}
\end{equation}

Table~\ref{tab:bskres} gives our results and Figure~\ref{fig:Kfplot} 
plots the results for $R_{B_sK}$ against $m_{\pi}$. We observe 
no significant $a$ dependence but some dependence on $m_{\pi}$, 
because the $K$ meson contains a valence $l$ quark.  
To fit the dependence of $R_{B_sK}$ as a function of $m_{\pi}$ and 
$a$ we use a similar fit form to that used earlier (eq.~(\ref{eq:fitform})). 
We drop the odd powers 
of $m_{\pi}$ from kinematic effects, since these now depend 
on $m_K$ which depends on $m_l$ and hence $m^2_{\pi}$. 
As for $B_s \rightarrow \eta_s$ we add a term allowing for mistuning of the 
$s$ quark mass as $h(m_{\eta_s}^2 - [0.6885\,\mathrm{GeV}]^2)$ with prior 
on h of 0.0(2). We also keep the chiral logarithm term from 
$B \rightarrow \pi$ but we expect the coefficient to be a lot smaller 
here so we allow a prior for the coefficient of 0(1). 
Again, for this quantity, we will only extract a value at the physical value of 
$m_{\pi}$ (and not $m_{\pi} \rightarrow 0$) and, 
since we have results very close to that point, 
the value is insensitive to the details of the fit. 

Our fit has a $\chi^2/{\mathrm{dof}}$ of 0.2 for 5 degrees of 
freedom and gives a result at physical $m_{\pi}$ for $R_{B_sK}$ of 
0.626(6) with error budget given in Table~\ref{tab:errs}. 
This can be converted into a value for 
$f_0(q^2_{max})$ using $f_{B_s}$ = 0.224(5) GeV~\cite{DowdallBdecay}
and $f_{K^+}$ = 0.1561 GeV and 
$(1+m_{K}/m_{B_s})$ = 1.0927~\cite{pdg}.  
We find 
\begin{equation}
\left. f_0(q^2_{max})\right|_{B_s\rightarrow K} = 0.822(8)(18) .
\label{eq:f0resK}
\end{equation} 
Here the first error is from $R_{B_sK}$ and the second from 
$f_{B_s}$. 

\section{Discussion}
\label{sec:discussion}

\begin{figure}
\begin{center}
\includegraphics[width=\hsize]{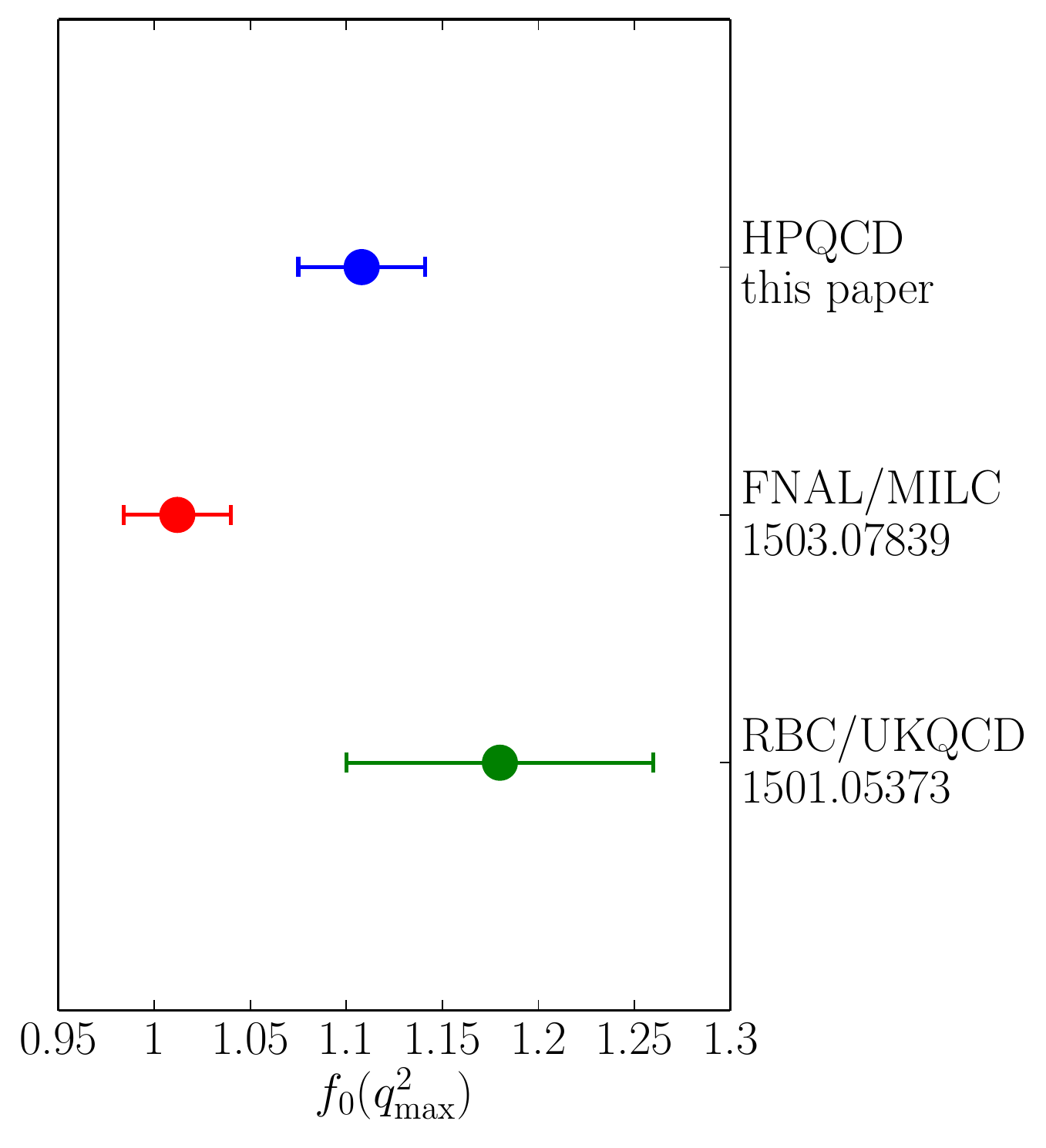}
\end{center}
\caption{ A comparison of results for $f_0(q^2_{max})$ 
for $B\rightarrow \pi l \nu$ decay. 
The top result is from this paper using gluon field configurations 
that include $u$, $d$, $s$ and $c$ quarks in the sea and include 
$u/d$ quarks at physical masses. 
The lower two results~\cite{FNALbtopi2015, RBCbtopi2015} use gluon field configurations that include 
$u$, $d$ and $s$ quarks in the sea and extrapolate to physical 
$m_{\pi}$ from heavier values. 
For all points statistical and systematic uncertainties have been added 
in quadrature. 
}
\label{fig:fcomp}
\end{figure}

We can compare our results for $f_0(q^2_{max})$ for $B \rightarrow \pi$ 
and $B_s \rightarrow K$ to existing full lattice QCD results 
where an extrapolation in $m_{\pi}$ to the physical point has been done 
(as well as a study of $f_+$ and $f_0$ as a function of $q^2$). 

Figure~\ref{fig:fcomp} shows this comparison for $B \rightarrow \pi$, 
comparing our result to that from Fermilab/MILC~\cite{FNALbtopi2015} and 
RBC/UKQCD~\cite{RBCbtopi2015}. Both of these calculations use a 
`clover' formalism for the $b$ quark, which has a nonrelativistic 
interpretation on coarse lattices seamlessly matching to a relativistic 
formulation as $a \rightarrow 0$. Details differ significantly in the two 
cases~\cite{ElKhadra:1996mp, Christ:2006us} with the RBC/UKQCD calculation tuning the clover term 
nonperturbatively and including $\mathcal{O}(\alpha_s a)$ 
current corrections to reduce discretisation errors. 
The Fermilab/MILC results couple 
their $b$ quark to an asqtad staggered quark on the MILC 2+1 asqtad 
configurations. They cover a range of lattice spacing values from 0.12 fm 
to 0.045 fm and have a minimum light quark mass corresponding to 
a value for $m_{\pi}$ of 180 MeV on 0.09 fm lattices. 
The RBC/UKQCD results use domain-wall light
quarks on  their 2+1 domain-wall configurations with two lattice 
spacing values (0.11 fm and 0.086 fm) 
and a minimum light quark mass corresponding to 
a value for $m_{\pi}$ of 290 MeV. 

As discussed above, our results have the advantage of 
including much lower values of $m_l$, down 
to values of physical $m_{\pi}$ (indeed our {\it maximum} 
light quark mass corresponds to $m_{\pi}$ of 305 MeV) as well as using 
a more highly-improved action (to reduce discretisation errors) 
for both the $b$ quark, the light quark, the current connecting 
them, and the gluon field. 

We see agreement between our result and that from RBC/UKQCD within 
their larger uncertainties. There is a 2$\sigma$ `tension' between our 
result and that of Fermilab/MILC where we have similar 3\% uncertainties. 
The Fermilab/MILC value for $f_0(q^2_{max})$ for 
$B^+ \rightarrow \pi^0$ is 1.012(28)~\cite{rzhou} to be compared to our 
result of 1.108(33) (eq.~\ref{eq:f0res2}) for the same case.  

For the processes $B_s \rightarrow K$ and $B_s \rightarrow \eta_s$ 
the extrapolation in $m_{\pi}$ is less of an issue and instead, for example, 
it becomes important to tune the $s$ quark mass accurately. 
For our $B_s \rightarrow K$ result we can compare 
to RBC/UKQCD as above~\cite{RBCbtopi2015} and also to an earlier 
HPQCD result~\cite{Bouchard:2014ypa} 
(which also gives $B_s \rightarrow \eta_s$ 
form factors). This latter result uses $\mathcal{O}(v^4)$ NRQCD for 
the $b$ quark with HISQ valence light quarks on 
MILC 2+1 asqtad lattices at two lattice spacing values 
(0.12 fm and 0.09 fm) and with a minimum light sea quark mass 
corresponding to $m_{\pi}$ = 280 MeV. 
In all cases we see agreement with our results, given 
in eqs~(\ref{eq:f0resetas}) and~(\ref{eq:f0resK}), within 
the larger $\mathcal{O}(5\%)$ uncertainties 
of~\cite{RBCbtopi2015, Bouchard:2014ypa}. For example~\cite{RBCbtopi2015} 
quotes a result for $f_0$ for $B_s \rightarrow K$ 
at $q^2 = 23.4 \mathrm{GeV}^2$ of 0.81(6) to be compared 
to our result at $q^2_{max}$ of 23.7 $\mathrm{GeV}^2$ of 
0.822(20) (eq.~(\ref{eq:f0resK})).   

In the comparison of $B/B_s$ semileptonic
form factors to experiment for extraction of $V_{ub}$ 
it should be emphasised that 
it is the $f_+$ form factor that is relevant for $\mu\nu$ or 
$e\nu$ final states, and not $f_0$. 
However, similar issues arise in both cases for the extrapolation 
to physical $m_{\pi}$ and so the tests above for $f_0$ are also 
relevant to $f_+$.

\section{Conclusions}
\label{sec:conclusions}

In this paper we have laid to rest a long-standing controversy over 
the relationship between the form factor $f_0$ at zero recoil in 
$B \rightarrow \pi$ decay and the ratio $f_B/f_{\pi}$ from lattice QCD results. 
We calculate the ratio of $f_0$ to $f_B/f_{\pi}$ directly and 
obtain particularly accurate results because: 
\begin{itemize}
\item our lattice $b$-light currents are accurate through $\alpha_s\Lambda/m_b$
and the renormalisation of the current cancels between $f_0$ and $f_B$.
\item we work with light quarks that have their physical mass as well 
as heavier masses to allow an extrapolation to $m_{\pi}=0$. This is 
well controlled because of the form the staggered quark chiral perturbation 
theory takes (see Appendix~\ref{appendix:schipt}). 
\item we use improved actions for our $b$ quarks, light quarks 
and gluon fields, so that discretisation errors are reduced below 
$\alpha_s a^2$. 
\end{itemize}
We find 
\begin{equation}
\left. \frac{f^{B \rightarrow \pi}_0(q^2_{max})}{f_B/f_{\pi}}\right|_{m_{\pi}=0} =  0.987(51)
\end{equation}
in agreement with the soft-pion theorem result of 1. 
This test adds confidence to our control of lattice systematic 
uncertainties now that we have reached values of the $\pi$ mass 
close to the physical point. 

We are then able to determine values of the $f_0$ form factor 
at zero recoil and for physical quark masses for 3 processes, 
obtaining the ratios defined in Section~\ref{sec:latt}: 
\begin{eqnarray}
R_{B\pi}(\mathrm{phys}) = 0.805(16) \nonumber \\
R_{B_sK}(\mathrm{phys}) = 0.626(6) \nonumber \\
R_{B_s\eta_s}(\mathrm{phys}) = 0.740(5) .
\end{eqnarray}
Numbers vary by 30\% as the quark content in initial and 
final state changes between $l$ and $s$, with corresponding 
changes in the value of $q^2_{\mathrm{max}}$. 
This allows us to extract:
\begin{eqnarray}
\left. f_0(q^2_{max})\right|_{B^+\rightarrow \pi^0} &=& 1.108(22)(24)\nonumber \\ 
\left. f_0(q^2_{max})\right|_{B_s\rightarrow K} &=& 0.822(8)(18) \nonumber \\
\left. f_0(q^2_{max})\right|_{B_s\rightarrow \eta_s} &=& 0.811(5)(16) .
\end{eqnarray}
These can act as calibration values for lattice QCD calculations 
working at heavier-than-physical light quark masses. 
The uncertainties we have reached in this first `second-generation' 
form-factor calculation 
are encouraging for the improvements that will be possible as we move 
away from zero recoil.

{\bf{Acknowledgements}} We are grateful to the MILC collaboration 
for the use of 
their configurations, to R. Horgan, C.Monahan and 
J. Shigemitsu for calculating the pieces needed for the current 
renormalisation used here, and to R. Zhou for useful discussions 
on the Fermilab/MILC results. 
Computing was done on the Darwin supercomputer at the University 
of Cambridge as part of STFC's DiRAC facility. 
We are grateful to the Darwin support staff for assistance. 
Funding for this work came from STFC, 
the Royal Society and the Wolfson Foundation. 

\appendix

\section{Why are staggered quarks so continuum-like?}
\label{appendix:schipt}
The $\slashed D$ operator in staggered-quark lattice
discretisations has $\mathcal{O}(a^2)$ lattice-spacing corrections that do not
vanish in the limit of zero quark mass~$m_q$. These are
due to taste-changing interactions and affect the zero modes of the
lattice Dirac operator $i\slashed{D}+m_q$ in the chiral limit, altering the
topological properties of the theory at non-zero 
lattice spacing~\cite{Follana:2004sz, Follana:2005km}. 
In principle, therefore, one should take
the continuum  limit~$a\to0$ before taking the chiral limit~$m_q\to0$; that is,
there will be a smallest quark mass for each lattice spacing below which 
non-physical effects will show up.

In practice, these non-physical effects have been small, especially with the 
HISQ discretisation and similar actions designed to suppress taste-changing 
interactions~\cite{tastechanging}. 
Such possibilities have therefore had minimal effect on analyses of 
meson decay constants and masses (see, for example,~\cite{Dowdallfkpi}). 
This might seem surprising because we are now working 
at realistic $u/d$ quark masses which are very small. 
Here we are also examining the limit as 
$m_{\pi} \rightarrow 0$ (based on results at non-zero 
values of $m_{\pi}$ and $a$).  

Chiral perturbation theory, and, in particular, staggered chiral perturbation
theory is a useful tool for analyzing such effects. 
The problem is then apparent
from the fact that the masses of different tastes of pion are split by 
taste-changing interactions of $\mathcal{O}(a^2\alpha_s^2, a^2\alpha_s^3)$, so
that only the Goldstone pion's mass vanishes when the quark mass goes to zero.
This affects the chiral logarithms in the theory since these typically involve 
averages over the masses of all tastes of pion 
in staggered chiral perturbation theory.
As long as taste-splittings are small compared to the Goldstone pion mass 
then the difference is purely a discretisation effect. 
However, in the opposite limit of $m_{\pi} \to0$, 
 $m_\pi^2\log(m_\pi^2)$ in the continuum theory is replaced 
by $a^2\log(a^2)$ in the staggered-quark theory. While such 
terms vanish as $a\to0$, one might still worry that
unusually large $a^2$ corrections
and corrections that are not analytic in~$a^2$ would make the approach to
the continuum limit hard to control in practice, as well as distorting 
the $m_{\pi}$ dependence.

In fact, as we show here, the leading non-analyticities in~$a^2$
cancel as $m_{u/d}\to0$ when using the HISQ discretisation at 
current lattice spacings, at least for such phenomenologically important
quantities as: $f_\pi$, $m_\pi^2$, $f_K$, $f_B$ and $f_0^{B \rightarrow \pi}$. 
Consequently HISQ quarks are more 
continuum-like in their $a$ and $m_{\pi}$ dependence 
than might have been expected, even for very small
$u/d$~masses. 

The non-analytic $a^2$~dependence coming from the normal chiral logarithms is 
cancelled by contributions from the `hairpin' diagrams in staggered 
chiral perturbation theory. We will show how this cancellation works in 
the next subsection. The cancellation occurs only for special values of 
the staggered chiral theory parameters 
$\delta_V^\prime$ and~$\delta_A^\prime$,
but at the same values for each of the physical quantities mention above.
Simulations show that the actual values are within 10\% or so of the
special values required for cancellation.

\subsection{Staggered Chiral Perturbation Theory analysis}

\begin{figure}
\begin{center}
\includegraphics[width=0.95\hsize]{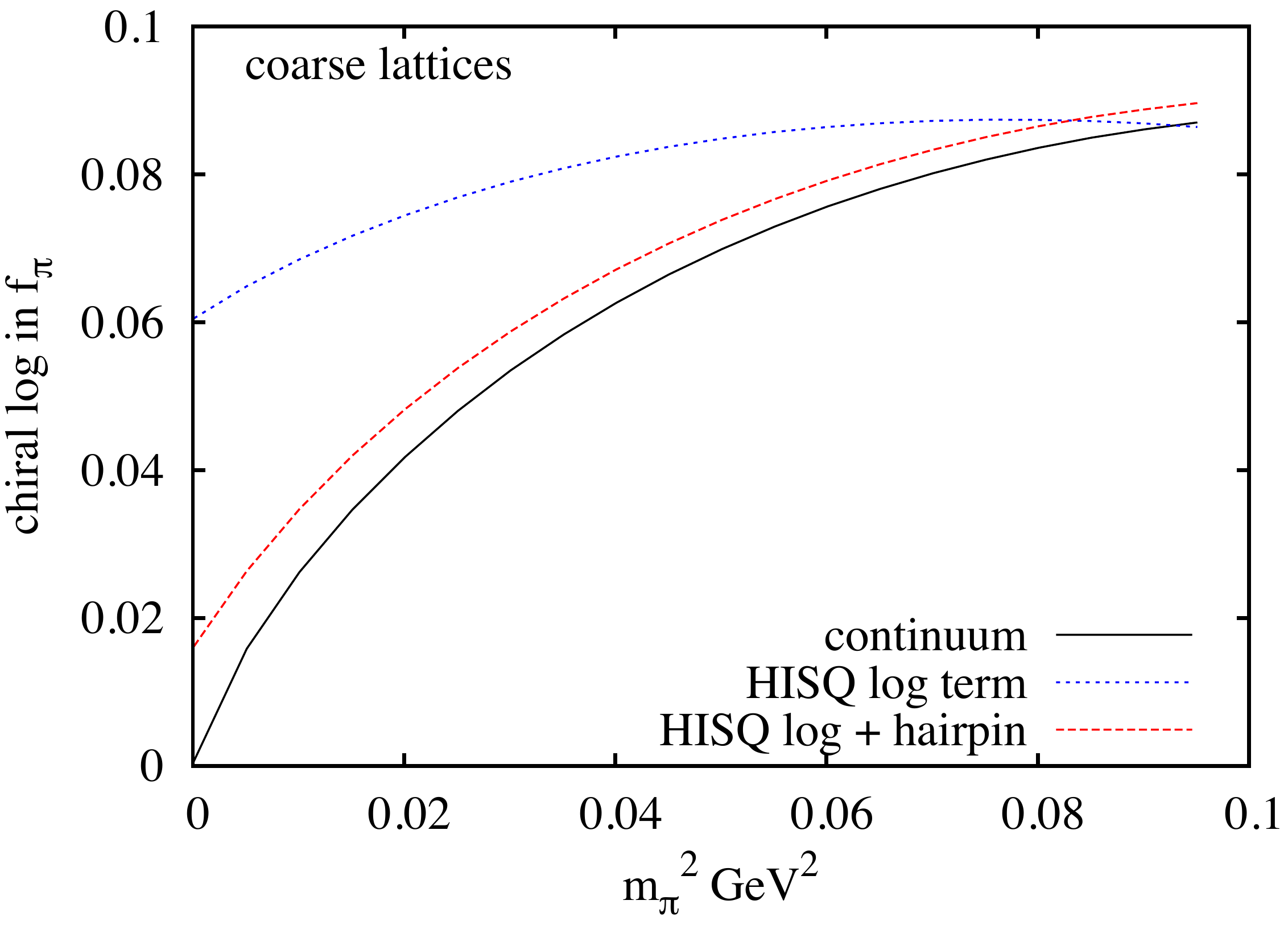}
\includegraphics[width=0.95\hsize]{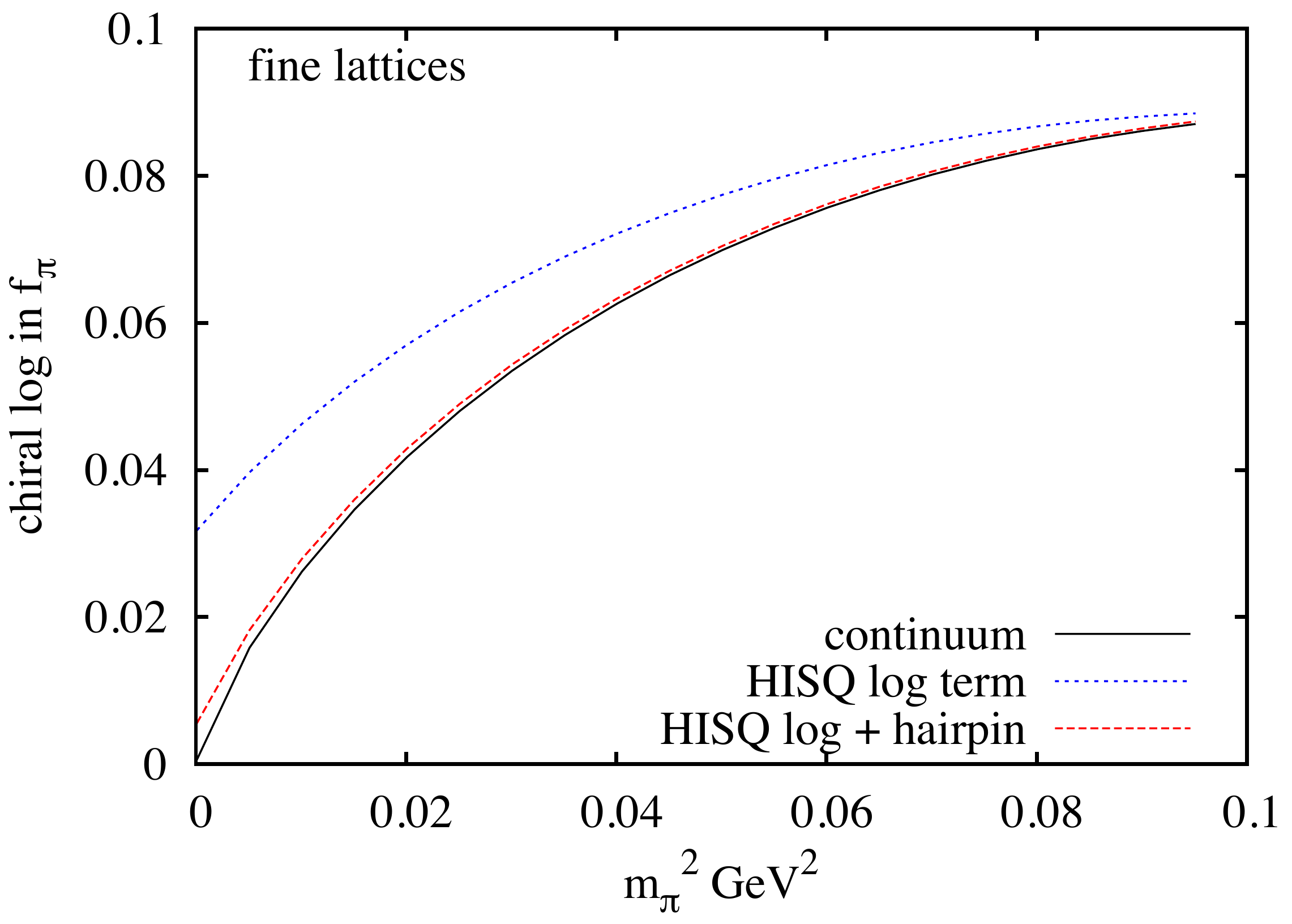}
\end{center}
\caption{ 
The chiral logarithm term in $m_{\pi}$ that appears in the chiral 
expansion of $f_{\pi}$ is shown as a function of $m_{\pi}$ 
in the continuum (solid black line) and compared to the results 
for the HISQ action from staggered chiral perturbation theory 
including logarithms and `hairpin' terms (red dashed line). The 
blue dotted lines shows results if only the `staggered' chiral 
logarithm is included. Results appropriate to our coarse lattices 
are shown in the upper plot, with results for the fine lattices 
in the lower plot. 
}
\label{fig:fpichiral}
\end{figure}

\begin{figure}
\begin{center}
\includegraphics[width=0.95\hsize]{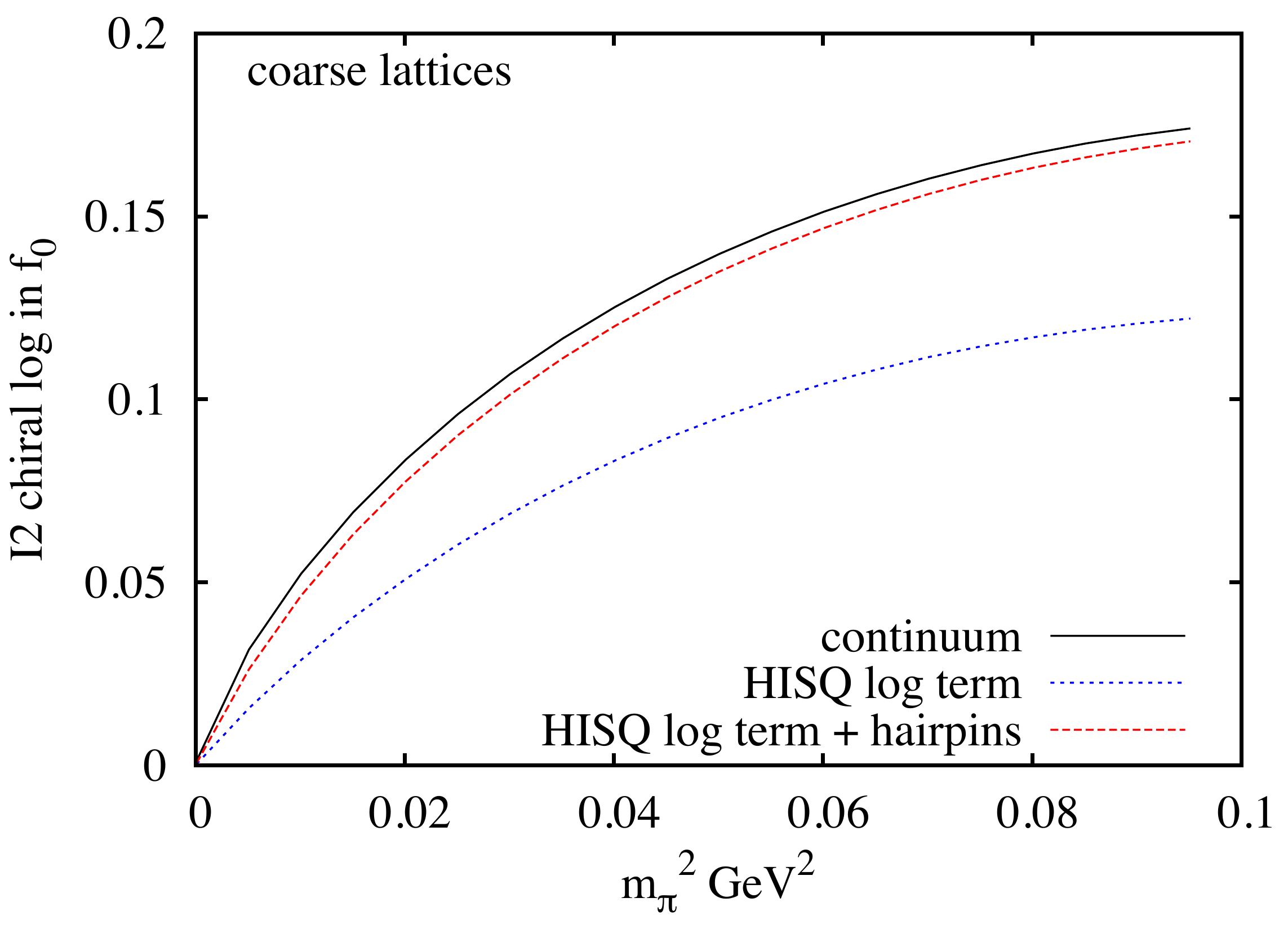}
\includegraphics[width=0.95\hsize]{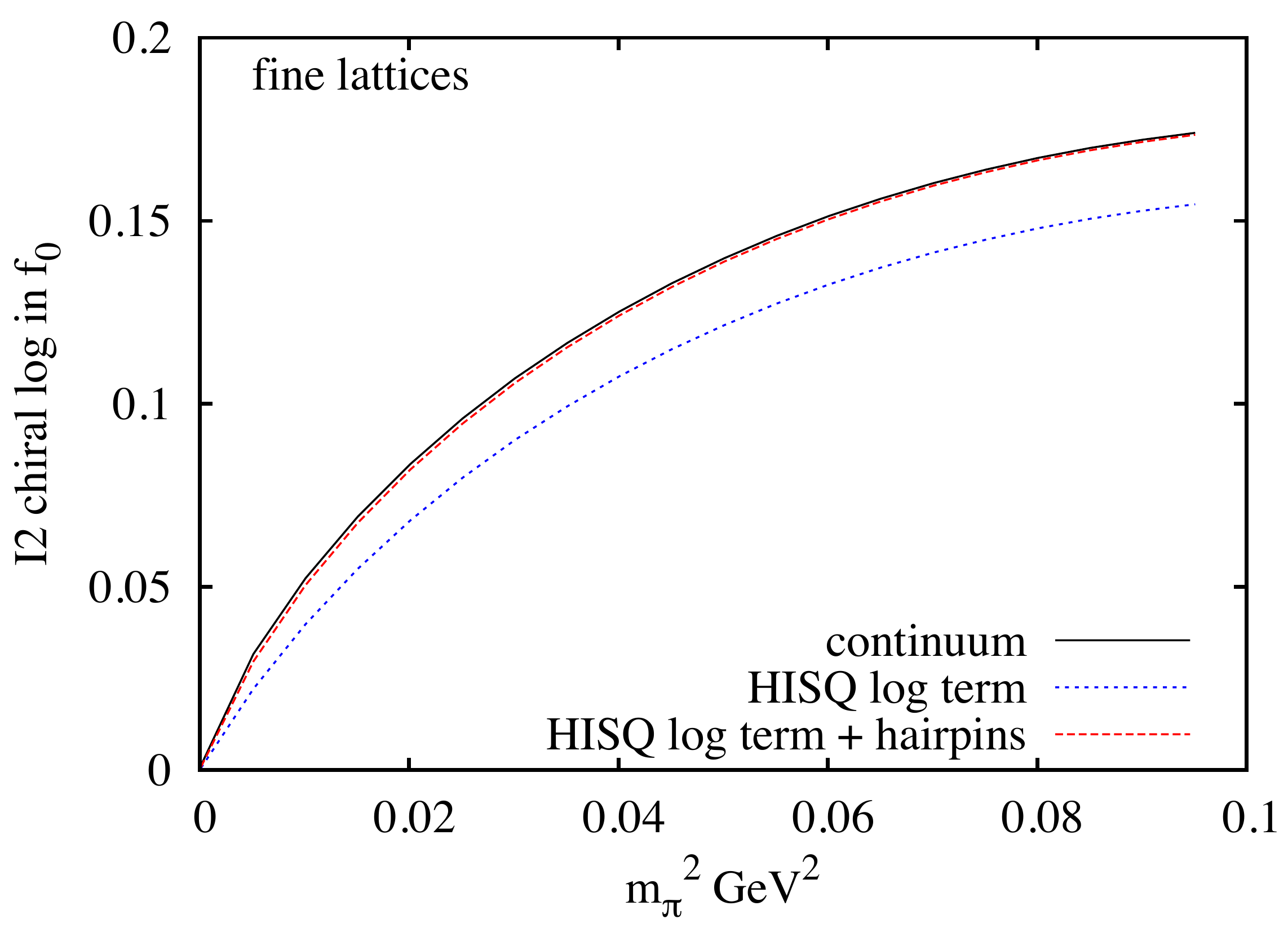}
\end{center}
\caption{ 
The chiral logarithm denoted $I2$ that appears in the chiral 
expansion of $f^{B \rightarrow \pi}_{0}(q^2_{max})$ 
is shown as a function of $m_{\pi}$ 
in the continuum (solid black line) and compared to the results 
for the HISQ action from staggered chiral perturbation theory 
including logarithms and `hairpin' terms (red dashed line). The 
blue dotted lines shows results if only the `staggered' chiral 
logarithm is included. Results appropriate to our coarse lattices 
are shown in the upper plot, with results for the fine lattices 
in the lower plot. 
}
\label{fig:f0I2}
\end{figure}

As examples, we first 
study $f_{\pi}$, $f_K$ and $f_{B}$ in the 2+1 full QCD case 
using formulae from~\cite{Aubin:2003uc} (eqs. 27 and 28) 
and~\cite{Aubin:2005aq} (eq. 105). We will evaluate here 
the leading behaviour in $a$ at Goldstone $m_{\pi}=0$ ($m_{u/d}=0$). The key terms 
are those that contain logarithms of masses of different tastes 
of pion - different tastes of kaon or other strange 
particle are irrelevant because their masses will be controlled 
by the $s$ quark mass. 

We implement a model for different tastes of pion in 
which the tastes are equally spaced. This is a very 
good approximation to what happens in simulations with 
the Highly Improved Staggered Quark action~\cite{milchisq} (and 
indeed the asqtad improved staggered quark action). 
This is an indication that the taste-breaking potential in 
the staggered chiral Lagrangian is dominated by one particular 
term~\cite{Lee:1999zxa, Aubin:2003mg, hisqdef}.
If we take the unit of splitting to be $\delta_t$ (proportional 
to $\alpha^2a^2$ or even $\alpha^3a^2$ in an improved formalism) 
then the Goldstone (G) pion 
mass will be zero  when $m_{u/d} = 0$ and the other tastes (axial vector, tensor, vector 
and singlet) will have squared masses 
according to:\\ 
$G : 0$ \\
$A : \delta_t$ \\
$T : 2\delta_t$ \\
$V : 3\delta_t$ \\
$I : 4\delta_t$. \\
The degeneracies of the different tastes from the Goldstone 
upwards in mass are: 1, 4, 6, 4, 1, making a total of 16 tastes. 

Then we can evaluate the chiral log terms that appear 
in the NLO term multiplying $1/(16\pi^2f^2)$ in staggered 
chiral perturbation theory at $m_{u/d}=0$. 
For $f_{\pi}$ this is simply:
\begin{eqnarray}
\label{eq:fpilog}
-\frac{1}{8}\sum_t m_{\pi,t}^2 \log{m_{\pi,t}^2} &=& -\frac{1}{8} (4\delta_t\log \delta_t + 12\delta_t\log 2\delta_t \nonumber \\ 
&& + 12\delta_t\log 3\delta_t + 4\delta_t\log 4\delta_t )   \nonumber \\
&=& -4\delta_t\log \delta_t + C\delta_t 
\end{eqnarray}
where $C$ is a constant. The $C\delta_t$ term is then a normal discretisation error 
which we will ignore here.  
Similarly for $f_K$ and $f_B$ (which has an overall extra 
factor of $1+3g_{B^*B\pi}^2$) the log term appears as: 
\begin{eqnarray}
\label{eq:fBlog}
&&-\frac{1}{16}\sum_t m_{\pi,t}^2 \log{m_{\pi,t}^2} + \frac{1}{4} m_{\pi, I}^2 \log{m_{\pi, I}^2} =  \nonumber \\
&& -\frac{1}{16} \left(4\delta_t\log \delta_t +  
 12\delta_t\log 2\delta_t + 12\delta_t\log 3\delta_t \right. \nonumber \\
&& + \left. 4\delta_t\log 4\delta_t \right)
+ \delta_t\log 4\delta_t  \nonumber \\
&=& -\delta_t\log \delta_t + C\delta_t  
\end{eqnarray}
Thus it is clear that the chiral logarithm terms on their own 
will give rise to potentially troublesome 
$\delta_t\log \delta_t \equiv a^2\log a^2$ terms (ignoring 
$\alpha_s$ factors in $\delta_t$ although in practice they make a significant 
numerical difference) as $m_{u/d}\rightarrow 0$.

In addition there are `hairpin' terms at NLO that come in a variety 
of forms. The key ones that can contain terms of the 
form $a^2\log a^2$ as $m_{u/d} \rightarrow 0$ 
are those multiplying logarithms of the 
mass of a taste 
of pion or a taste of $\eta$ (but not the singlet). 
For $f_{\pi}$ there are two such hairpin terms that we 
reproduce here in the axial taste case (there is also a vector 
taste version of them)~\cite{Aubin:2003uc}. Term 1 is : 
\begin{equation}
\label{eq:hpin1}
T1 =  \frac{(m^2_{S_A}-m^2_{\eta_A})}{(m^2_{\pi_A}-m^2_{\eta_A})(m^2_{\eta^{\prime}_A}-m^2_{\eta_A})}\ell(m_{\eta_A})
\end{equation}
where $\ell(m)=m^2\log m^2$ and this term appears with 
coefficient $2\delta_A^{\prime}$ which is itself $\mathcal{O}(a^2)$ 
(we absorb the $a^2$ into $\delta_A^{\prime}$ here). 

To evaluate $T1$ we use~\cite{Aubin:2003mg}:
\begin{eqnarray}
m^2_{\eta_A} &=& \frac{1}{2}(m^2_{\pi_A} + m^2_{S_A} + 0.75\delta_A^{\prime} - Z) \nonumber \\
m^2_{\eta^{\prime}_A} &=& \frac{1}{2}(m^2_{\pi_A} + m^2_{S_A} + 0.75\delta_A^{\prime} + Z) 
\end{eqnarray}
\begin{eqnarray}
&& Z = \nonumber \\
& & \sqrt{(m^2_{S_A}-m^2_{\pi_A})^2-0.5\delta_A^{\prime}(m^2_{S_A}-m^2_{\pi_A})+9(\delta_A^{\prime})^2/16} \nonumber \\
\end{eqnarray}
where $m^2_{S_A}$ is the axial taste of the $s\overline{s}$ pseudoscalar which 
we take to have the same taste-splittings as the pion. Again this is a very 
good approximation in the HISQ case~\cite{milchisq}. 

We can evaluate $Z$ to first order in $\delta_A^{\prime}$ at $m_{u/d}=0$ as:
\begin{equation}
Z = (m^2_{S_A}-m^2_{\pi_A}) - \delta_A^{\prime}/4
\end{equation}
and then the $\eta$ and $\eta^{\prime}$ masses follow: 
\begin{eqnarray}
m^2_{\eta_A} &=& m^2_{\pi_A} +\delta_A^{\prime}/2 \\
m^2_{\eta^{\prime}_A} &=& m^2_{S_A} +\delta_A^{\prime}/4 \nonumber 
\end{eqnarray}
Then we can evaluate each of the mass differences in $T1$:
\begin{eqnarray}
m^2_{\eta^{\prime}_A}-m^2_{\eta_A} &=& Z \\
&=& m^2_{S_G} - \delta_A^{\prime}/4  \nonumber
\end{eqnarray}
\begin{equation}
m^2_{S_A}-m^2_{\eta_A} = m^2_{S_G} - \delta_A^{\prime}/2 
\end{equation}
\begin{equation}
\label{eq:etaA}
m^2_{\pi_A}-m^2_{\eta_A} = - \delta_A^{\prime}/2. 
\end{equation}
This final mass difference, appearing in the denominator 
of $T1$ looks dangerous but the whole term, as discussed 
above, is multiplied by $\delta_A^{\prime}$. 

Combining all the mass differences we have: 
\begin{eqnarray}
2\delta_A^{\prime} T1 &=& 2\delta_A^{\prime} \frac{m^2_{S_G}}{(-\delta_A^{\prime}/2) m^2_{S_G}}\ell(m_{\eta_A})\nonumber \\
&=& -4 m^2_{\eta_A}\log m^2_{\eta_A} 
\end{eqnarray}
to leading order in $\delta_A^{\prime}$. However $\delta_A^{\prime}$ still appears inside 
$m^2_{\eta_A}$. 

We approach $T2$ similarly. 
\begin{equation}
\label{eq:hpin2}
T2 =  \frac{(m^2_{S_A}-m^2_{\pi_A})}{(m^2_{\eta_A}-m^2_{\pi_A})(m^2_{\eta^{\prime}_A}-m^2_{\pi_A})}\ell(m_{\pi_A})
\end{equation}
From above we have the mass differences: 
\begin{eqnarray}
m^2_{\eta^{\prime}_A}-m^2_{\pi_A} &=& m^2_{S_G} +\delta_A^{\prime}/4  \\
m^2_{\eta_A}-m^2_{\pi_A} &=& + \delta_A^{\prime}/2  \nonumber \\
m^2_{S_A}-m^2_{\pi_A} &=& m^2_{S_G}  \nonumber
\end{eqnarray}
Then 
\begin{eqnarray}
2\delta_A^{\prime} T2 &=& 2\delta_A^{\prime} \frac{m^2_{S_G}}{(\delta_A^{\prime}/2) m^2_{S_G}}\ell(m_{\pi_A})\nonumber \\
&=& 4 m^2_{\pi_A}\log m^2_{\pi_A} 
\end{eqnarray}

Adding $T1$ and $T2$, as required for the chiral 
expansion of $f_{\pi}$ gives:
\begin{eqnarray}
&& 4 (m^2_{\pi_A}\log m^2_{\pi_A} - m^2_{\eta_A}\log m^2_{\eta_A}) \\
&=&
  4(\delta_t\log \delta_t - (\delta_t+\delta_A^{\prime}/2)\log(\delta_t+\delta_A^{\prime}/2) ) \nonumber 
\end{eqnarray}

Since $\delta_A^{\prime}$ also arises from staggered-quark 
taste effects 
we can assume that it is some multiple of 
the unit of taste-splitting, $\delta_t$. If we write it as a multiple, 
$x_A$, 
of the largest taste-splitting (between the taste-singlet and 
the Goldstone), then $\delta_A^{\prime}=4x_A\delta_t$. 
The sum of $T1$ and $T2$ becomes 
\begin{eqnarray}
 4(\delta_t\log \delta_t &-& (1+2x_A)\delta_t\log(1+2x_A)\delta_t)  \\
&=& 4(1-(1+2x_A))\delta_t \log \delta_t + C\delta_t \nonumber \\
&=& -8x_A\delta_t\log \delta_t  \nonumber
\end{eqnarray}
neglecting terms of $\mathcal{O}(\delta_t)$. An equivalent result 
would be obtained for the vector hairpin terms.  

Combining the chiral log for $f_{\pi}$ with the $\delta_A^{\prime}$ 
and $\delta_V^{\prime}$ hairpin 
terms then gives:
\begin{eqnarray}
-4\delta_t \log \delta_t &-& 8(x_A+x_V)\delta_t \log \delta_t = \\
&& -4(1+2[x_A+x_V])\delta_t\log \delta_t  \nonumber
\end{eqnarray}
We see that $x_A+x_V=-0.5$ is a special value where the $\delta_t\log \delta_t$ (i.e. 
$a^2\log a^2$) terms cancel between the chiral logarithms 
and the hairpins.  In fact this is the value that is obtained 
from staggered chiral perturbation theory fits to $f_{\pi}$ and 
$f_K$. Such fits in~\cite{Dowdallfkpi} give 
\begin{eqnarray}
\label{eq:deltaval}
x_A &=& -0.31(4) \\
x_V &=& -0.25(6) \nonumber
\end{eqnarray} 
so that $\delta_A^{\prime} \approx \delta_V^{\prime} \approx -\delta_t$. 
Then the axial taste of $\eta$ has a mass between that of 
the Goldstone and axial tastes of pion from eq.~(\ref{eq:etaA}). 

The impact of this is made clear in Figure~\ref{fig:fpichiral} 
in which we plot the chiral logarithm term as a 
function of the (Goldstone) $m_{\pi}$. The solid black 
curve shows the continuum 
form $-2(m_{\pi}^2/\Lambda_{\chi}^2)\log (m_{\pi}^2/\mu^2)$ 
with $\Lambda_{\chi}=4\pi f_{\pi}$ (for $f_{\pi}$ = 130 MeV) 
and $\mu=0.568$ GeV. 
For comparison we show the result from simply replacing 
the continuum chiral logarithm by an average over 
chiral logarithms for each taste of pion (the term 
analysed in eq.~(\ref{eq:fpilog})). We take 
equally-spaced masses for the tastes with splittings 
appropriate to coarse and fine lattices from~\cite{milchisq}. 
The distortion of the continuum chiral logarithm 
from taste-splitting discretisation effects is 
clear, particularly on the coarse lattices at small 
values of $m_{\pi}$. In contrast, when the complete 
staggered chiral perturbation theory expression is 
taken, including the hairpin corrections
discussed above, the taste-splitting discretisation effects 
are almost eliminated and the match with the continuum chiral 
perturbation theory behaviour is restored. 
The curves in Figure~\ref{fig:fpichiral} used 
$\delta_A^{\prime}=\delta_V^{\prime}=-\delta_t$ with
$\delta_t = 0.022 \,\mathrm{GeV}^2$ on coarse lattices 
and $\delta_t = 0.007 \,\mathrm{GeV}^2$ on fine lattices. 
The impact of taste-changing discretisation effects on 
the chiral perturbation theory is
then very small, even on the coarse lattices. 

To understand why the full staggered chiral perturbation 
theory looks so continuum-like across the full range of $m_{\pi}$ 
values (and not just as $m_{\pi} \rightarrow 0$) 
we can consider another limit in which $\delta_t < m_{\pi}^2$ with 
$0 < m_{\pi}^2 < m_{\eta_s}^2$. This region is in reasonable 
correspondence with the right-hand side of our plots 
in Figure~\ref{fig:fpichiral}. 
Then the full staggered chiral perturbation theory gives 
the continuum chiral logarithms as well as terms that 
behave as $\delta_t \log m_{\pi}^2$ and 
$m_{\pi}^2 \log(1+n\delta_t/m_{\pi}^2) \equiv n\delta_t$. 
Both of these terms look like regular discretisation errors and 
are not problematic. However, once again, both of these terms 
cancel between the staggered chiral 
logarithms and the hairpin terms for the case above where 
$\delta_A^{\prime}=\delta_V^{\prime}=-\delta_t$.  
This explains why there is such good correspondence, with only very small 
discretisation errors visible, between 
the full staggered chiral perturbation theory and the continuum 
chiral logarithm for $f_{\pi}$ across the full 
range of $m_{\pi}^2$ values in Figure~\ref{fig:fpichiral}. 

The staggered chiral perturbation theory for 
$f_K$ and $f_B$ is similar to that for $f_{\pi}$~\cite{Aubin:2003uc, Aubin:2005aq}. 
The hairpin contributions that 
can give $a^2\log a^2$ terms as $m_{\pi}^2 \rightarrow 0$ 
in these cases are identical to the 
$T1$ and $T2$ above for $f_{\pi}$ 
(eqs.~(\ref{eq:hpin1}) and~(\ref{eq:hpin2})). They come with 
a coefficient of $1/2$ instead of 2, however (again 
$f_B$ has the additional multiplier of $1+3g^2_{B^*B\pi}$). 
This is exactly the overall factor of $1/4$ needed 
to cancel the $\delta_t\log \delta_t$ piece of the chiral logarithm 
(eq.~(\ref{eq:fBlog}))
when $x_A+x_V =-0.5$ as for $f_{\pi}$.  
So again in this case staggered chiral perturbation 
theory behaves much more benignly than might be 
expected in its approach to the continuum limit. 
This will also be true, as for $f_{\pi}$, across 
our range of $m_{\pi}$ values. 

The staggered chiral perturbation 
theory for $m_{\pi}$ (in terms of $m_{u/d}$) 
is somewhat different from the examples above~\cite{Aubin:2003mg}. 
The `chiral log' is simply $\ell(m_{\pi_I})$. 
For $m_{u/d}=0$ this becomes $4\delta_t\log 4\delta_t$ multiplying 
the standard factor of $1/(16\pi^2f^2)$. 

The `hairpin' terms are also somewhat different. 
For the axial case, keeping only the pieces that 
can give rise to $\delta_t\log \delta_t$ terms,  we have: 
\begin{eqnarray}
T3 &=& -4\ell(m_{\pi_A}) \nonumber \\ 
&-& \frac{2\delta_A^{\prime}}{m^2_{\eta^{\prime}_A} - m^2_{\eta_A}}\frac{m^2_{S_A}-m^2_{\eta_A}}{m^2_{\pi_A}-m^2_{\eta_A}}\ell(m_{\eta_A})
\end{eqnarray} 

Using mass-splittings from above this becomes 
\begin{eqnarray}
T3 &=& -4\ell(m_{\pi_A}) - 2\delta_A^{\prime} \frac{m^2_{S_G}}{m^2_{S_G}(-\delta_A^{\prime}/2)}\ell(m_{\eta_A}) \nonumber \\
&=& -4 \ell(m_{\pi_A}) +4 \ell(m_{\eta_A}) 
\end{eqnarray}
This shows how the $\log m_{\pi}$ factors in the 
$a \rightarrow 0$ limit cancel 
between the two halves of T3. 
To obtain the $a^2\log a^2$ pieces as $m_{u/d} \rightarrow 0$ 
we need to take the Goldstone $m_{\pi}$ to zero. 
Then we have, for the axial case: 
\begin{eqnarray}
T3 &=& -4\delta_t \log \delta_t + 4 (1+2x_A) \delta_t \log(1+2x_A)\delta_t  \nonumber \\
&=& 8x_A\delta_t \log \delta_t 
\end{eqnarray}
Again, adding in the vector taste piece, we find 
that the value $x_A+x_V=-0.5$ cancels 
the $\delta_t\log \delta_t$ behaviour from 
the chiral log term above. 
So the same values for the 
$\delta_A^{\prime}$ and $\delta_V^{\prime}$ coefficients 
also give rise here to a cancellation of 
taste-splitting artefacts in the staggered chiral perturbation theory. 
It is also true, as before, that this cancellation can be demonstrated 
explicitly for the $m_{\pi}^2 >0$ case when $\delta_t < m_{\pi}^2$ 
and in practice numerically it works across the whole $m_{\pi}$ 
range used here. 

For the form-factor $f_0^{B \rightarrow \pi}$ discussed here 
we use the staggered chiral perturbation theory given 
in~\cite{Aubin:2007mc} (eq.67). Not surprisingly a set of 
terms appear there containing logarithms of $m^2_{\pi,t}$ 
that are the same as those appearing in the staggered chiral 
perturbation theory for $f_{\pi}$ and $f_B$. 
Our analysis above then demonstrates that these terms 
(combining chiral logarithms and hairpins)
behave as continuum chiral perturbation theory. 
In fact these terms (denoted by $I_1$ in~\cite{Aubin:2007mc}) 
cancel between $f_0$ and $f_B/f_{\pi}$ in the ratio 
$R_{B\pi}$ that we consider here. 

There are additional terms in $f_0$ 
denoted $I_2$ in~\cite{Aubin:2007mc}
that also behave as chiral logarithms. 
They have the form $(v.p)^2\log(m^2_{\pi,t}/\mu^2)$ 
where $v.p$ is the dot product of $B$ meson
velocity and $\pi$ 4-momentum. For a $B$ meson 
at rest and a Goldstone $\pi$ in the final 
state this gives $E^2_{\pi,G} \log(m^2_{\pi,t}/\mu^2)$, 
reducing to $m^2_{\pi,G} \log(m^2_{\pi,t}/\mu^2)$ at 
zero recoil. These are the chiral logarithm terms 
that survive in the ratio $R_{B\pi}$ and that 
we include in our fit of eq.~(\ref{eq:fitform}). 
Such terms only contain $\delta_t$ 
inside the logarithm and so cannot give 
rise to $\delta_t \log \delta_t$ terms.  
The approach to $m_{\pi} \equiv m_{\pi,G}=0$ for $R_{B\pi}$ 
is therefore the same 
as in continuum chiral perturbation theory even 
at non-zero lattice spacing.  
Discretisation effects could arise from the $m^2_{\pi,t}$ 
inside the logarithm but these can be shown to cancel 
in the $\delta_t < m_{\pi}^2$ case for the same 
$\delta_A^{\prime}$ and $\delta_V^{\prime}$ values 
(in eq.~(\ref{eq:deltaval}). 
Once again the numerical result, shown in Fig.~\ref{fig:f0I2}, gives 
continuum-like behaviour for the staggered chiral 
perturbation across the full range of $m_{\pi}$ values. 
We nevertheless allow for possible $m_{\pi}$-dependent 
discretisation errors 
in our chiral fit, as discussed in Section~\ref{sec:results}. 

\bibliography{bpipaper}

\end{document}